\documentclass[12pt,reqno]{article}
\usepackage[utf8]{inputenc}
\usepackage{a4wide}
\usepackage{amssymb,amsmath,amsthm,newlfont,enumerate}
\usepackage{appendix}
\usepackage{dsfont}
\usepackage{amsfonts}
\usepackage{amssymb}
\usepackage{amsbsy}
\usepackage[dvips]{graphicx}
\usepackage{epstopdf}
\usepackage{psfrag}
\usepackage[hang,center]{caption}
\usepackage{verbatim} 
\usepackage{float}
\usepackage{cite}
\usepackage[colorlinks=true,linkcolor=blue,urlcolor=blue]{hyperref}
\usepackage{color}
\usepackage[normalem]{ulem}
\usepackage{fancyhdr}
\usepackage{bm}
\usepackage{xcolor}
\usepackage{empheq}
\usepackage{placeins}
\usepackage[all,cmtip]{xy}
\usepackage{nccmath}

\def\RR{\mathbb{R}}

\def\cA{\mathcal A}

\def\cR{\mathcal R}
\def\cQ{\mathcal Q}

\def\cT{\mathcal T}

\def\cL{\mathcal L}

\def\bfy{\mathbf{y}}
\def\bfZ{\mathbf{Z}}
\def\bfz{\mathbf{z}}
\def\bfG{\mathbf{G}}

\let\eps\varepsilon

\let\tn\textnormal
\let\vphi\varphi
\let\pa\partial
\let\para\parallel
\let\wh\widehat

\def\be{\begin{equation}}
\def\ee{\end{equation}}
\def\bes{\begin{equation*}}
\def\ees{\end{equation*}}

\DeclareMathOperator{\arctantwo}{arctan2}

\newcommand{\parfra}[2]{\frac{\partial #1}{\partial #2}}

\newcommand{\dt}[1]{\frac{\mathrm d #1}{\mathrm dt}}
\newcommand{\dtpr}[1]{\frac{\mathrm d #1}{\mathrm dt'}}
\newcommand{\ds}[1]{\frac{\mathrm d #1}{\mathrm ds}}

\newcommand{\gavg}[1]{\langle #1 \rangle}

\theoremstyle{definition}
\newtheorem{definition}{Definition}
\newtheorem{assumption}{Assumption}
\newtheorem{remark}{Remark}
\newtheorem*{example*}{Example}

\theoremstyle{plain}
\newtheorem{lemma}{Lemma}
\newtheorem{prop}{Proposition}
\newtheorem{theorem}{Theorem}
\newtheorem*{theorem*}{Theorem}

\title{Gyrokinetics from variational averaging:\\ existence and error bounds}

\author{
Stefan Possanner\footnote{stefan.possanner@ma.tum.de}
\\[5mm]
{\small {\it  Technische Universität München, Zentrum Mathematik, Boltzmannstraße 3, 85748 
Garching}}
\\[4mm]
{\small {\it  Max-Planck-Institut für Plasmaphysik, Boltzmannstraße 2, 85748 Garching }}
}

\date{}

\begin{document}
\maketitle

\begin{abstract}
The gyrokinetic paradigm in the long wavelength regime is reviewed from the perspective of 
variational averaging (VA). The VA-method represents a third pillar for averaging kinetic 
equations with  highly-oscillatory characteristics, besides classical averaging or Chapman-Enskog 
expansions. VA operates on the level of the Lagrangian 
function and preserves the Hamiltonian structure of the characteristics at all orders. We discuss 
the methodology of VA in detail by means of charged-particle motion in a strong 
magnetic field. The application of VA to a broader class of highly-oscillatory problems can be 
envisioned. For the charged particle, we prove the existence of a coordinate map in phase space 
that leads to a gyrokinetic Lagrangian at any order of the expansion, for general external fields. 
We compute this map up to third order, independent of 
the electromagnetic gauge. Moreover, an error bound for the solution of the derived 
gyrokinetic equation with respect to the solution of the Vlasov equation is provided, allowing to 
estimate the quality of the VA-approximation in this particular case.
\end{abstract}

\bigskip


\bigskip

{\bf AMS subject classifiaction:} 34C29, 35Q75, 70H09, 78A35, 82D10.

\section{Introduction}

Charged particles in a strong magnetic field are spiraling around their ``center of motion'', 
the 
gyro-center (GY). The stronger the magnetic field, the smaller the gyro-radius $\rho_\tn{s}$ and the 
larger the gyro-frequency  $\omega_\tn{c}$ of the spiraling motion; the 
charged-particle dynamics is usually a multiscale problem. Low-frequency 
($\omega/\omega_\tn{c} \ll 1$) and large 
scale ($x/\rho_\tn{s} \gg1$) phenomena become of interest for instance in space physics or in 
magnetically confined 
fusion devices \cite{Krommes2012gyrokinetic}. The modeling of these 
phenomena relies on averaging the gyro-motion, leading to reduced dynamics for the GY. This is a 
prototypical example of perturbation theory in (nearly-) periodic dynamical systems 
\cite{Sanders2007}, with important consequences for practical applications. The GY-dynamics are 
appealing for numerical simulations of 
large ensembles of charged particles, giving rise to gyrokinetic equations. Such models are 
implemented in many state-of-the art computer codes for plasma turbulence simulations
\cite{ELMFIRE,EUTERPE,GEM,GENE,GTC,GYRO,GYSELA,ORB5}.

Given a large ensemble of charged particles, the kinetic (Vlasov) equation for the phase 
space distribution $f$ of these particles reads
\be \label{Vlasov}
 \dt{} f(x_\eps(t),v_\eps(t),t) = 0\,.
\ee
The solution $f$ is constant along the characteristics $x_\eps(t) \in \RR^3$, $v_\eps(t) \in\RR^3$, 
which satisfy Newton's equations of motion under the Lorentz force. The high-dimensionality of the 
problem makes its numerical solution rather arduous. Moreover, in strong magnetic fields the 
characteristics $x_\eps(t)$, $v_\eps(t)$ are highly-oscillatory with a period $\eps\ll 1$, i.e. 
$x_\eps(t) = x(t/\eps)$ leading 
to severe time step restrictions in numerical solvers. Instead of following the exact trajectory, 
reduced dynamics for the GY have proven to be useful in numerical experiments. Gyrokinetic equations 
have been derived on three different levels:
\begin{enumerate}
 \item directly from the kinetic equation \eqref{Vlasov} via a Chapman-Enskog expansion of the 
solution, $f = f_0 + \eps\,f_1 + \ldots$,
 \item from the characteristics by averaging the dynamical system $x_\eps(t)$, $v_\eps(t)$,
 \item on the level of the Lagrangian via ``variational averaging'' (VA).
\end{enumerate}
In this work we shall focus on the third approach, variational averaging. VA places the emphasis on 
the Hamiltonian structure of the dynamical system, which is preserved in the process. The formal 
theory of VA has been developed in the early eighties in the plasma physics community 
\cite{Little1979,Little1981,Little1982,Little1983,Brizard1989,Hahm1988}. A pioneering work for 
averaging nearly-periodic Hamiltonian systems has been given by Kruskal \cite{Kruskal}. Up to 
now mathematically rigorous results for VA are lacking, which is surprising considering its 
importance for numerical plasma physics. In this work we shall close this mathematical gap and 
establish
several cornerstones of variational GY-theory:

\begin{itemize}
\item existence of a GY-transformation leading to reduced dynamics;
\item gauge-invariance;
\item definition of a gyrokinetic equation;
\item strong error estimate for the gyrokinetic solution with respect to $f$, solution of 
\eqref{Vlasov}.
\end{itemize}

We are able to prove existence with a new ansatz for the GY-transformation as a finite power series 
in $\eps$, algebraic in the generating functions, in contrast to the usual Lie-transform approach, 
which relies on operator exponentials of Poisson brackets. Prerequisites for understanding existing 
formal VA-theories \cite{Hahm1988,Hahm1996,Brizard_rev2007,Parra2011,Tronko2017} include a 
firm knowledge about exterior calculus, differential forms and Lie transforms, with rare 
exceptions \cite{Little1983,Scott2017}. Our theory does not 
rely on these concepts and is thus more accessible for non-specualists. The long 
wavelength regime is considered, hence the inclusion of finite-Larmor radius effects postponed to a 
future work. We stress the non-uniqueness of transformations leading to GY-Lagrangians, which is 
overlooked in the existing VA-theories. A new GY-transformation is presented which leads to simpler 
equations of motion; this is possible due to the freedom of ``unloading'' complicated terms into 
the 
transformation (the generating functions), rather than keeping them in the Lagrangian.

The methodology of VA is carefully developed in this work. The concept of the ``tangent map'' 
between two coordinate representations of a manifolds's tangent bundle is introduced in detail. We 
then shift the focus to a particular class of Lagrangian functions of the form \eqref{intro:L}, 
linear in the tangent vectors. The VA-theory developed here could in principle be applied to a 
large class of highly-oscillatory problems, formulated in terms of this generic 
Lagrangian. The charged particle is a prototypical example and treated in detail.

Historically, the first approach towards reduced GY-models stems from averaging Newton's 
equation of motion for the charged particle \cite{Jackson,Northrop1963}. Assuming a uniform static 
magnetic field, these can be solved exactly to yield the spiraling motion around the straight field 
lines. In this case the GY is well-defined and its trajectory follows a magnetic field line. Adding 
a static perpendicular 
electric field gives rise to a drift across field lines, but the GY is still well-defined. The 
problem complicates when the fields have curvature (non-homogeneous case). 
In this case several new drifts appear, for instance the curvature drift and the grad-$B$ drift 
\cite{Hazeltine}. On top of that, the GY is no longer well-defined: the center of the spiral cannot 
be computed in closed form, its location only approximated by an infinite series. In the 
non-homogeneous case the GY-dynamics are thus truncated dynamics (perturbation theory). 
For the validity of the theory it is thus essential to control the error that arises 
from truncation.

VA is based on a variational principle from which the equations of motion can be derived. Since 
the variational principle is coordinate independent it is particularly suited for averaging, which 
is nothing else than a change of coordinates, with a minimum amount of algebra. 
VA has the advantage that the Hamiltonian structure of the particle dynamics is not destroyed in 
the process. This leads in 
particular to conservation of a truncated energy and to conservation of a truncated phase space 
volume, which are easily identified. These and other conservation properties related to the 
Hamiltonian structure are beneficial for stability and accuracy of long-time numerical simulations.

Let us briefly mention some mathematical results on averaged particle dynamics in strong magnetic 
fields, not related to VA: Fr\'enod and Sonnendr\"ucker 
\cite{FrenodSonnen1998,FrenodSonnen2001} 
use two-scale convergence to establish limit models of the Vlasov-Poisson system in strong magnetic 
fields. The asymptotic behavior of the Vlasov-Maxwell system in strong magnetic fields has been 
considered by Bostan \cite{Bostan2007,Bostan2010,Bostan2010transport,Bostan2016}, relying on energy 
methods or averaging techniques. The transition from Vlasov to gyrokinetic equations has also been 
studied for 
example in \cite{Golse1999,Han2010,Filbet2016asymp}. Stroboscopic averaging is applied to the 
GY-problem in \cite{Chartier2012,Chartier2016}. A WKB-based approach with emphasis on gyro-gauge 
has been presented in \cite{Tronci2016}. 

The article is organized as follows: in the preliminary section \ref{sec:prelim} we clarify some 
notation in \ref{subsec:notation} and introduce 
the equations of motion and their normalization in \ref{sec:model}. In \ref{subsec:var} we discuss 
the corresponding variational formulation; the notion of a Lagrangian 
function defined on the tangent bundle of the underlying manifold is presented in detail. We 
formulate the guiding-center problem as well as the full problem with electromagnetic external 
fields in the extended phase space. In section \ref{sec:finite} we introduce the method of VA, which 
is based on the concept of the tangent map. Our new ansatz for the algebraic GY-transformation is 
stated here.
The main results are collected in section \ref{sec:results}, which is split into three subsections:
in the preliminary part we define the guiding-center Lagrangian, equivalence of 
Lagrangians and the 
gyro-average operation and state the existence of solutions for the charge-particle dynamics in 
~\ref{subsec:prelim}. Section \ref{subsec:exist} contains the existence results for the 
GY-transformation in Theorem \ref{thm0}. Explicit expressions for the GY-transformation, the 
corresponding GY-Hamiltonian and the generalized magnetic moment are given in 
section~\ref{sec:express}. Finally, a gyrokinetic equation is defined in section~\ref{subsec:res2}; 
its strong solution is compared to the solution of the Vlasov equation in Theorem \ref{thm}.
Proofs that require a lot of algebra have been put into section \ref{sec:proofs}. We summarize 
the article and discuss future perspectives in section \ref{sec:concl}.

\section{Preliminaries}  \label{sec:prelim}

\subsection{Notation} \label{subsec:notation}

The vector product in $\RR^3$ is denoted by '$\times$'. The symbol '$\nabla$' denotes the usual 
gradient operator in $\RR^3$, hence $\nabla=(\pa_{x_1},\pa_{x_2},\pa_{x_3})^\top$. For 
a vector field $A:\RR^3\to\RR^3$, $A=(A_1,A_2,A_3)^\top$, we write $\nabla \times A$ to denote the 
curl-operator. Given a map $\tau:\RR^n\to\RR^n$, the Jacobian is denoted by $D\tau$, i.e. 
$(D\tau)_{i,j} = \pa \tau_i / \pa x_j$. For $n=3$ we denote the transpose Jacobian by
$$
   \nabla A := \Big( 
\parfra{A_j}{x_i} \Big)_{1\leq i,j\leq3} = \Big(\parfra{A}{x}\Big)^\top = (DA)^\top  \,.
$$
The dot '$\cdot$' denotes the scalar product in Euclidean space; it is also used to denote 
matrix-vector multiplication in $\RR^n$. For $b\in\RR^3$ for example
$$
 (b \cdot \nabla) A = b \cdot \nabla A = \nabla A^\top \cdot b\,.
$$

\subsection{Equations of motion and scaling}  \label{sec:model}

Newton's equation of motion for a non-relativistic charged particle in an electromagnetic field 
can be written as 
\be \label{Newton}
 \dt x = v\,,\qquad\quad \dt v = \frac{e}{m} \Big[ v \times B(x,t)  + E(x,t) \Big]\,.
\ee
Here, $x$ stands for the particle position, $v$ its velocity, $e$ the particle's charge, $m$ its 
mass and $B$ and $E$ denote external magnetic and electric fields. The 
right-hand-side in the equation for $v$ is the Lorentz force, hence gravitational and other effects 
are neglected. Our first task is to formulate Newton's equation of motion in dimensionless form. 
For example, we write the solution $x$ as $x(t) = \hat x \,x'(t')$, where $\hat x$ denotes the 
characteristic size (scale or unit) of the particle position and $x'$ is a dimensionless function 
of $t' = t/\hat t$, the time in units of $\hat t$. The characteristic size $\hat x$ could be for 
instance the diameter of our domain of study and $\hat t = \hat\omega^{-1}$, where $\hat\omega$ 
characterizes the frequency domain of interest. Similarly, $B(x,t) = \hat B\,B'(x',t')$ for the 
fields. 
Hence,
\be \label{Newton:2}
 \dtpr x' = \frac{\hat v}{\hat x\,\hat\omega }\,v'\,,\qquad\quad \dtpr v' = \frac{e\,\hat 
B}{m\,\hat \omega} \Big[ v' \times B'(x',t')  + \frac{\hat E}{\hat v\,\hat B}\, E'(x',t') \Big]\,.
\ee
The characteristic cyclotron frequency of the problem is $\hat \omega_\tn{c} = e\,\hat B/m$. We 
simplify via
\be
 \tn{a)}\quad \hat v = \hat x\,\hat \omega\,,\qquad\qquad \tn{b)}\quad \eps:= \frac{\hat 
\omega}{\hat \omega_\tn{c}} \,,\qquad\qquad \tn{c)}\quad \eps_\delta := \frac{\hat E}{\hat 
v\,\hat B} \,.
\ee
In assumption a) we relate the velocity scale $\hat v$ to the chosen time- and space scales $\hat 
\omega^{-1}$ and $\hat x$. In b) we introduce a first parameter $\eps$; if $\eps \ll 1$ one enters 
the low-frequency regime, which means that the frequency of interest $\hat 
\omega$ is much smaller than the cyclotron frequency~$\hat \omega_\tn{c}$. A second parameter 
$\eps_\delta$ is introduced in c); it represents the ratio of the $E$$\times$$B$-velocity 
to the characteristic velocity $\hat v$. This parameter will also appear in the magnetic field, 
which we assume to be composed of two parts: 
\be \label{assume:B}
 B(x,t) = B_0(x) + \eps_\delta\, B_1(x,t)\,,
\ee
a so-called ``guide field'' $B_0$, which is static and non-homogeneous and a dynamical part $B_1$ 
with amplitude $\eps_\delta$. Thus $\eps_\delta$ 
signifies the amplitude of the dynamical fields $E/v$ and $B_1$ with respect to the static guide 
field $B_0$.
We introduce a third parameter $\eps_B$ which measures the degree of inhomogeneity of the guide 
field ($||\cdot||$ is some matrix norm): 
\be \label{def:epsB}
 \eps_B := \hat x\,\frac{||\nabla B_0||}{|B_0|}\,.
\ee
Two cases of $\eps_B$ shall be addressed in this paper: $\eps_B = 1$, which signifies that the 
guide field variations are on the scale $\hat x$, and $\eps_B = \eps$ which corresponds to less 
important variations of the guide field. 

Let us now insert the above definitions of $\eps$-parameters into Newton's equations 
\eqref{Newton:2} and omit the primes to obtain 
\be \label{Newton:3}
 \dt x = \,v\,,\qquad\qquad \dt v = \frac{1}{\eps}\, v \times \Big[ B_0(\eps_B x) + 
\eps_\delta\,B_1(x,t)\Big]  + \frac{\eps_\delta}{\eps}\, E(x,t) \,.
\ee
 Two orderings shall be addressed in this work:
\be \label{scalings}
 1)\quad \eps_\delta = \eps\,,\quad \eps_B = 1 \,,\qquad\qquad 2)\quad \eps_\delta = \eps_B = 
\eps\,.
\ee
Ordering 1) is rarely discussed in GY-theory, whereas case 2) is called the ``maximal 
ordering'' \cite{Brizard_rev2007}. We point out that the ordering $\eps_B = \eps$ is implemented in 
all the aforementioned gyrokinetic models used for computer simulations, because of its 
relative simplicity with respect to the case $\eps_B = 1$ at the second order of expansion (see 
below).

\subsection{Variational formulation} \label{subsec:var}

\subsubsection{Problem statement}
Under the scaling assumptions from the previous section, the initial-value problem (IVP) we 
consider reads
\be \label{ivp}
\left\{
 \begin{aligned}
  \dt x &= v\,, &&\qquad x(t_0) = x_0\,,
  \\[2mm]
  \dt v &= \frac{v \times B_0(\eps_B x)}{\eps} + v \times B_1(x,t) + E(x,t) \,, &&\qquad v(t_0) = 
v_0\,.
 \end{aligned}
\right.
\ee
Here, we assume $x,x_0\in \Omega_x\subset \RR^3$, $v,v_0 \in \Omega_v \subset \RR^3$ with 
$\tn{dim}(\Omega_x) = \tn{dim}(\Omega_v) = 3$, ${\Omega =\Omega_x 
\times \Omega_v}$ open and bounded and $0< \eps 
\leq \eps_\tn{max}$. For $\eps\ll 1$ system \eqref{ivp} represents a multi-scale problem with a 
fast, nearly-periodic motion around $B_0$. Classical averaging \cite{Sanders2007} can be applied to 
extract reduced dynamics free of the fast scale. However, system \eqref{ivp} is also rich in 
structure, a so-called Hamiltonian system. In order 
to see the structure we need to study its variational formulation.

\subsubsection{Lagrangian functions} \label{subsec:Lag}

The variational formulation of \eqref{ivp} is based on a Lagrangian function, simply 
called the ``Lagrangian''. Lagrangians are defined on the tangent bundle of the underlying 
manifold, which in our study is the phase space, and map into the real numbers. We 
shall clarify this notion in more detail. 

Let  $M\subset \RR^n$ denote an open subset of Euclidean space $\RR^n$ with 
points $\mathfrak m \in M$, described by a 
single coordinate chart $\vphi: U\subset \RR^n \to 
M$, $q \mapsto \mathfrak m$ ($M$ is thus an $n$-dimensional differentiable manifold). $q$ are 
called coordinates of $M$ under the chart~$\vphi$. The tangent space at point $\mathfrak m \in M$, 
denoted by $T M_{\mathfrak m}$, is the space af all vectors originating from $\mathfrak m$, hence 
$T M_{\mathfrak m} = \RR^n$. More precisley $T M_{\mathfrak m}$ contains equivalence classes of 
curves 
through $\mathfrak m$, two curves being equivalent when they are tangent to each other at 
$\mathfrak m$ 
\cite{Marsden,Arnold}. We denote by $\xi \in T M_{\mathfrak m}$ an element of the tangent space 
at~$\mathfrak m$. 

Coordinates for $\xi \in T M_{\mathfrak m}$ can be constructed from the chart 
$\vphi$ as follows: for an open interval $I\subset\RR$ let $c:I\to U$ 
denote a curve in the coordinate space $U$ with $ c(0) = q$; then $\vphi( c)$ is a curve passing 
through $\mathfrak m$ at $t=0$ on the manifold $M$. The tangent at $t=0$ in the coordinate space is 
$\dot q := 
\dt{} c (0)$; in the tangent space $TM_{\mathfrak m}$ the tangent is
\be
 \xi = \dt{\vphi( c(t))} \Big|_{t=0} = \sum_j \parfra{\vphi}{q_j}\Big|_{ c(0)} \dt{ c_j 
(0)} = D\vphi(q) \cdot \dot q \,.
\ee
Since this holds true for any curve $ c$ passing through $q$ at $t=0$ we deduce that any $\xi 
\in TM_{\mathfrak m}$ can be written in the form $D\vphi(q) \cdot v$ for some $v \in \RR^n$. It 
follows that the tangent space $TM_{\mathfrak m}$ is the image of the Jacobian 
$D\vphi(q)$; a basis of $TM_{\mathfrak m}$ is thus given by the columns of $D\vphi(q)$, which we 
denote by $\pa_j := \parfra{\vphi}{q_j}$ (covariant basis) \cite{Frankel}. The coefficients of a 
tangent vector 
$\xi$ in this basis are denoted by $\dot q$, hence
\be \label{xi:dotq}
 \xi = D\vphi(q) \cdot \dot q = \sum_j \dot q_j\, \pa_j\,.
\ee
The union of all 
tangent spaces ``attached'' to $M$ at points~$\mathfrak m$ is called the tangent bundle $T M$; its 
elements are tangent vectors. The chart $\vphi$ induces coordinates in the tangent bundle which we 
denote by $(q,\dot q)$. The first coordinate $q\mapsto \mathfrak m$ identifies the tangent space 
and the second coordinate $\dot q \mapsto \xi$ identifies an element in that particular tangent 
space. 

Another useful object is the dual to the tangent space $TM_{\mathfrak m}$, called the cotangent 
space $T^*M_{\mathfrak m}$. Its elements are covectors or linear forms $\gamma: TM_{\mathfrak m} 
\to \RR$, mapping tangents into the real numbers. The chart $\vphi$ induces a basis also in the 
cotangent space: given the basis vectors $\pa_j$ of the tangent space, the dual basis $\tn d_i \in 
T^*M_{\mathfrak m}$ is defined by the property $\tn d_i(\pa_j) = \delta_{ij}$, where $\delta_{ij}$ 
is the Kronecker delta. Since for the Jacobains we have $D\vphi^{-1} D\vphi = I_n$ where 
$I_n$ is the identiy matrix, we deduce that the lines of $D\vphi^{-1}$ are the sought dual basis, 
thus $\tn d_i := \nabla \vphi_i^{-1}$ (contravariant basis). Denoting the components of $\gamma$ in 
this basis by 
$\gamma_i$ we have, for general $\gamma \in T^*M_{\mathfrak m}$ and $\xi \in TM_{\mathfrak m}$,
\be \label{gammaxi}
 \gamma(\xi) = \sum_i \gamma_i\,\tn d_i(\xi) = \sum_{ij} \gamma_i\,\dot q_j\, \tn d_i(\pa_j) = 
 \sum_{ij} \gamma_i\,\dot q_j\, \delta_{ij} = \gamma \cdot \dot q\,.
\ee
Hence the natural pairing between elements of the tangent space $T M_{\mathfrak m}$ and 
elements of its dual $T^*M_{\mathfrak m}$ can be written as a scalar product in $\RR^n$ with 
respect to the bases induced by the chart $\vphi$. We shall use this convenient notation throughout 
this work. 

The union of all cotangent spaces at points $\mathfrak m \in M$ is called the cotangent bundle and 
denoted by $T^*M$. The chart $\vphi$ induces coordinates in the cotangent bundle: an element is 
identified via $\gamma(q)$, where $q\mapsto \mathfrak m$ identifies the 
cotangent space (dual to the tangent space at $\mathfrak m$) and the ``vector'' $\gamma$ holds the 
components of the linear form in that particular cotangent space, such that the duality pairing can 
be written as the scalar product \eqref{gammaxi}.

We are now equipped to define a Lagrangian function on the tangent bundle of the manifold $M$.
We shall consider dynamical systems defined by a particular class of Lagrangians  $L: T M \to \RR$ 
which, in local coordinates $(q,\dot q)$ defined by some chart $\vphi:U\subset \RR^n\to M$, can 
be written as
\be \label{intro:L}
 L(q,\dot q) = \gamma(q) \cdot \dot q - H(q)\,.
\ee
Here, $H: M \to \RR$ is called the Hamiltonian and $\gamma \in T^*M$ is the symplectic form, in 
the sense of \eqref{gammaxi}. We will now discuss how the charged-particle problem \eqref{ivp} can 
be deduced from such a Lagrangian by a variational principle.

\subsubsection{The action principle}

Given the Lagrangian \eqref{intro:L} the dynamics follow from a variational principle on curves in 
the coordinate space 
$U$. Let us denote such curves by $q(s)$, or more precisely by $q:I\to U$ for some open
interval $I\subset \RR$. Let us further define the following functional on the space of curves,
\be \label{action}
 \cA[q] := \int_I L\Big(q(s),\ds{} q(s) \Big)\,ds\,.
\ee
The variational (action) principle $\delta \cA/\delta q = 0$ yields the Euler-Lagrange equations 
\be \label{EuLag}
 \parfra{L}{q} - \ds{} \parfra{L}{\dot q} = 0\,,
\ee
which, for $L$ given by \eqref{intro:L}, become
\be \label{sympl}
 \omega \cdot \ds q  = \parfra{H}{q}\,,
\ee
where $\omega := (D\gamma)^T - D\gamma$ is called the Lagrange matrix. We assume that $\omega$ is 
invertible on $U$ and write $J:=\omega^{-1}$. Then system \eqref{sympl} can be written as
\be \label{sympl:PB}
 \ds q = \{ q, H \}\,,
\ee
where $\{G,H\}:= \pa G/\pa q \cdot J \cdot \pa H/\pa q $ denotes the Poisson bracket, defined for
 differentiable functions $G,H:U\to \RR$. The bracket is bilinear, 
anti-symmetric and satisfies the Jacobi identity
$$
\{F,\{G,H\}\} + \{H,\{F,G\}\} + \{G,\{H,F\}\} = 0\,.
$$
Systems of the form \eqref{sympl} where $\omega$ is invertible are called non-canonical symplectic 
systems, which belong to the larger class of Hamiltonian systems. It is an immediate consequence of 
\eqref{sympl:PB} that $H(q)$ is a constant of the 
motion, ${\ds{} H = 0}$. Moreover, it can be shown that the flow of \eqref{sympl:PB} conserves 
the phase space volume $\sqrt{\det \omega}$, computed from the determinant of the Lagrange matrix 
$\omega$. Other constants of the motion are the so-called Casimirs and the momentum maps 
\cite{Marsden,Arnold}. Exact conservation of these invariants on the discrete level leads to 
improved long-time stability and accuracy of numerical schemes. An example of such a symplectic 
integrator is the well-known St\"ormer-Verlet scheme \cite{Hairer_geom}.

\subsubsection{The guiding-center problem}

If the dynamical fields $E$ and $B_1$ in \eqref{ivp} are zero, the problem of averaging reduces to 
the so-called guiding-center (GC) problem. In this case the system \eqref{ivp} is autonomous and a 
Lagrangian of the generic form \eqref{intro:L} can be formulated in the coordinate space $U= 
\Omega$; it reads
\be \label{L_a}
 L_\tn{a} = \Big( v +  \frac{A_0(x)}{\eps} \Big) \cdot \dot x - \frac{|v|^2}{2} \,.
\ee
Here, $A_0$ is the vector potential related to the guide field via $B_0 = \nabla \times A_0$. In 
terms of the generic form \eqref{intro:L} we have 
\be
 \gamma = \gamma_\tn{a} = \Big( v +  \frac{A_0(x)}{\eps}, 0,0,0 \Big)\,,\qquad\qquad H = H_\tn{a} 
= \frac{|v|^2}{2}\,.
\ee
The velocity components of the symplectic form are zero. It can be easily checked that the 
Euler-Lagrange equations corresponding to $L_\tn{a}$ yield the equations \eqref{ivp} without 
dynamical fields. Moreover, we know that this system is non-canonical symplectic because its 
Lagrange matrix is invertible. The kinetic energy $H_\tn{a}$ is conserved during the motion.

Variational averaging of the Lagrangian \eqref{L_a} has been studied extensively on the formal 
level; the first rigorous results are presented in this work. A review can be found 
in \cite{BrizardCary2009}. Higher-order computations of the asymptotic GC-expansion have recently 
been reported \cite{Brizard_Natalia2015,Burby}. The computations in this paper will reproduce the 
standard GC-results up to second order in the GC-Hamiltonian and GC-symplectic form. The 
first-order GC-Lagrangian is defined in \eqref{def:Lgc}.

\subsubsection{Full problem with dynamical fields}

In case that the dynamical fields $E$ and/or $B_1$ are not zero the system \eqref{ivp} is 
non-autonomous. It becomes an autonomous system in the extended coordinate space ${U= 
\Omega\times \RR^2}$ with coordinates $q = (x,v,t,w)$. Here, the time $t$ and the 
energy $w$ are dependent variables and the independent variable is denoted by $s$. Symplectic form 
and Hamiltonian are introduced as
\be
 \gamma_\tn{ext} := \Big( \gamma_\tn{a},0,0\Big) + \Big( A_1(x,t),0,0,0,-w,0 \Big)\,, \qquad\qquad 
H_\tn{ext} := H_\tn{a} + \phi(x,t) - w\,. 
\ee
Here, the dynamical electromagnetic potentials $A_1$ and $\phi$ are such that
\be \label{def:A}
  B_1 = \nabla \times A_1\,,\qquad\qquad E = -\nabla \phi - \parfra{A_1}{t}\,.
\ee
The Lagrangian is of the generic form \eqref{intro:L} and reads
\be \label{Lext}
 L = \Big( v + \frac{A_0(x)}{\eps} + A_1(x,t) \Big) \cdot \dot x - w\,\dot t - 
\frac{|v|^2}{2} - \phi(x,t) + w\,.
\ee
The corresponding Lagrange matrix is invertible and the system is non-canonical symplectic with 
conserved energy $H_\tn{ext}$. The Euler-Lagrange equation for $w$ automatically yields ${\ds{} t = 
1}$ and thus $t = s$. The charged-particle 
dynamics are found to occur on the hyper-surface $H_\tn{ext} = 0$ of the extended coordinate 
space. For simplicity during variational averaging we directly impose $H_\tn{ext} = 0$ which means 
$w =  |v|^2/2 + \phi =: H$; this leads to the ``extended Lagrangian'' \cite{Arnold}
\be \label{def:LPC}
  L_\tn{I}:=L\,\Big|_{w = H} = \Big( v + \frac{A_0(x)}{\eps} + A_1(x,t) \Big) \cdot \dot x  - 
H\,\dot t  \,,
\ee
where the coordinate space is $\Omega_\tn{I} := 
\Omega \times \RR$ with elements $q=(x,v,t)$. The Lagrangian \eqref{def:LPC} is written as $L_\tn{I}(q,\dot q) = 
\gamma_\tn{I}(q)  \cdot \dot q$, where  
\be \label{def:PCform}
  \gamma_\tn{I} := \Big( v + \frac{A_0(x)}{\eps} + A_1(x,t),0,0,0,-H \Big)
\ee
is the well-known Poincar\'e-Cartan form; it is the starting point for any gyro-averaging theory 
in the variational framework.

\subsection{Change of coordinates}  \label{sec:finite}

\subsubsection{What is variational averaging?}
The aim of variational GY-theory is to preserve the symplectic structure of the charged-particle 
dynamics \eqref{ivp}, manifested by a Poisson bracket \eqref{sympl:PB}, when averaging the 
fast scale due to the $v\times B_0$ motion. The structure originates from the generic form of the 
Lagrangian~\eqref{Lext}. Hence, 
averaging directly on the level of the Lagrangian while keeping the generic form \eqref{intro:L} is 
the favorable strategy, as outlined in \cite{Little1983}. ``Averaging'' in this context can be 
defined by the following steps:
\begin{enumerate}
 \item Identify a fast variable, the gyro-angle, that changes on the time scale $\eps$ due 
to the $v\times B_0$ motion. This is done by a ``preliminary map'' in the extended 
Lagrangian~\eqref{def:LPC}.
 \item Find a change of coordinates in phase space that decouples the fast motion of the gyro-angle 
from the remaining equations on the slow scale. Suppose $\alpha$ denotes the fast 
variable, then the coordinate map should eliminate $\alpha$ from the Lagrangian at 
successive orders in $\eps$, up to the desired order $\eps^N$. 
 \item The ``decoupling'' is then accomplished by truncating the new Lagrangian at order~$N$, which 
means neglecting terms of order $\eps^{N+1}$:
\be \label{intro:Lgy}
 L^\eps = \underbrace{\frac{L_{-1}}{\eps} + L_0 + \eps\,L_1 + \ldots + 
\eps^N\,L_N}_{=: L_\tn{gy}^{(N)}} + \,\eps^{N+1}\,L_{N+1} +  \ldots\,.
\ee
 \item The ``decoupled'' equations of motion are the Euler-Lagrange equations stemming 
from the truncated Lagrangian $L_\tn{gy}^{(N)}$. They feature the slow variables which, by 
construction, can be computed independently of the fast variable $\alpha$. The term ``averaged 
dynamics'' refers to the dynamics of the slow variables. Moreover, the Euler-Lagrange equation 
\eqref{EuLag} for $\alpha$ yields 
\be \label{conserve}
 \ds{} \parfra{L_\tn{gy}^{(N)}}{\dot \alpha} = 0\,,
\ee
which states the conservation of the ``generalized magnetic moment'' $\wh\mu:=\pa L_\tn{gy}^{(N)} 
/\pa \dot \alpha$.
\end{enumerate}
We identify three fundamental questions related to the above approach:
\begin{itemize}
 \item Under what premise does a coordinate map leading to \eqref{intro:Lgy} exist?
 \item How does the truncation error in the Lagrangian translate to errors in the 
equations of motion?
 \item In what way can the averaged equations be used to derive a gyrokinetic equation?
\end{itemize}
These questions, among others, shall be addressed in the course of this work. The main tool for 
variational averaging is the ``tangent map'', which allows us to transform Lagrangians defined on 
tangent bundles; it is introduced next.

\subsubsection{The tangent map}

Let $M\subset \RR^n$ with $\mathfrak m \in M$ denote the manifold introduced in section 
\ref{subsec:Lag}, described by the single coordinate chart ${\vphi:U\subset\RR^n \to M}$, $q\mapsto 
\mathfrak m$. Suppose $\psi:V\subset\RR^n \to M$ stands for a different chart describing the same 
manifold $M$ in the coordinates $l\mapsto \mathfrak m$. Then the map $\tau: V\to U$, $l\mapsto q$ 
given by $q = \tau(l) = \vphi^{-1} \circ \psi(l)$ defines a change of coordinates on the manifold 
$M$. The map $\tau$ is one-to-one and differentiable with differentiable inverse, hence a 
diffeomorphism. Its Jacobian is $D\tau = D\vphi^{-1} D\psi$.

The transformation law for elements $\xi$ of the tangent space $TM_{\mathfrak m}$ is 
straightforward: from \eqref{xi:dotq} we have
\be
 \xi = D\vphi(q) \cdot \dot q = D\psi(l) \cdot \dot l\,.
\ee
The components $\dot  q$ can thus be expressed in terms of the components $\dot l$ via
\be 
\dot q = D\vphi(\tau(l))^{-1} D\psi(l) \cdot \dot l = D\tau(l) \cdot \dot l\,.
\ee
\begin{definition}({\bf Tangent map}.)
 Given a change of coordinates $\tau:l\mapsto q$ on the manifold~$M$, the 
associated ``tangent map'' $T\tau: (l,\dot l)\mapsto (q,\dot q)$ relating two coordinate systems of 
the tangent bundle $TM$ is defined by 
\be \label{tangent}
(q,\dot q) = T\tau(l,\dot l) := (\tau(l),D\tau (l) \cdot \dot l)\,.
\ee
\end{definition}

The tangent map is the principal tool for the theory of variational 
averaging presented in this work. It will be used to transform the extended Lagrangian 
\eqref{def:LPC} from the coordinates $q\in \Omega_\tn{I}$ to new coordinates $l \in V$:
\be \label{map:LI}
 L_\tn{I}(q,\dot q) = \gamma_\tn{I}(q) \cdot \dot q = \underbrace{\gamma_\tn{I}(\tau(l)) \cdot 
D\tau(l)}_{=:\wh \gamma_\tn{I}(l)} \cdot \dot l = \wh \gamma_\tn{I}(l) \cdot \dot l\,.
\ee
Here, we almost accidentally uncovered the transformation law of covectors (elements of the 
cotangent space), $\gamma_\tn{I} \circ \tau  = D\tau^{-T}\, \wh \gamma_\tn{I}$. Variational 
averaging is built on the fact that in \eqref{map:LI} the generic form of the extended Lagrangian 
$L_\tn{I}$ is preserved under the tangent map. Moreover, from the transformation law of cotangents 
we can deduce that the new Lagrange matrix $\wh \omega = (D \wh\gamma_\tn{I})^T - D 
\wh\gamma_\tn{I}$ is invertible, and hence the symplectic structure preserved.

\subsubsection{Preliminary transformation}

We apply a preliminary coordinate map to the extended Lagrangian $L_\tn{I}: T\Omega_\tn{I} \to \RR$ 
from~\eqref{def:LPC} for the purpose of identifying the fast variable (gyro-angle), which is then 
subjected to 
averaging. We start from a local, orthonormal basis 
$(e_1(x),e_2(x),b_0(x))$ that satisfies $b_0 = e_1 \times e_2$, $e_1 = e_2\times 
b_0$, $e_2 =b_0 \times e_1$ such that $b_0\cdot e_1 \times e_2 =1$ and the basis is 
right-handed. New velocity coordinates are introduced as
\be \label{transf}
\begin{aligned}
 v_\parallel &:= v\cdot b_0(x)\,,  
 \\[4.5mm]
 v_\perp &:= |b_0(x) \times v \times b_0(x)| = |v - v\cdot b_0(x) b_0(x)|\,, 
 \\[2mm]
 \theta &:= -\arctantwo\left( 
\frac{v\cdot e_2(x)}{v\cdot e_1(x)}\right)\,,
\end{aligned}
\ee
such that $v = v_\parallel b_0 + v_\perp c_0$, where $c_0 := e_1(x) \cos 
\theta - e_2(x)\sin\theta$.
Together with the unit vector $a_0 := e_1(x) \sin \theta + e_2(x) \cos\theta$, the triple 
$(a_0,b_0,c_0)$ is an orthonormal basis of $\Omega_v$ at each $x\in\Omega_x$. Moreover, one has the 
identities
\be \label{a_to_v}
 b_0 \times v = v_\perp a_0\,,\qquad\quad b_0 \times v\times b_0 = v_\perp c_0\,.
\ee
Now let $\Omega_\tn{I}'$ 
denote the extended phase space with velocity coordinates \eqref{transf}, i.e. for $q' \in 
\Omega_\tn{I}'$ we have $q' = (x,v_\para,v_\perp,\theta,t)$. The preliminary map is thus 
\be \label{prelim:tau}
\tau':  \Omega_\tn{I}' \to  \Omega_\tn{I}\,, \qquad q'\mapsto q\,,
\ee
defined by 
\be \label{prelim}
x = x\,,\qquad\quad  v = v_\parallel b_0(x) + v_\perp c_0(x,\theta)\,,\qquad\quad t = t\,,
\ee
with Jacobian determinant $-v_\perp$. The transformed Lagrangian $L'$ is obtained from 
\eqref{def:LPC} by inserting \eqref{prelim},
\be \label{L:X}
L'(q',\dot q') = \Big[ v_\para b_0(x) + v_\perp c_0(x,\theta) + 
\frac{A_0(x)}{\eps} + A_1(x,t)   \Big] \cdot \dot x - \Big[ \frac{v_\para^2}{2} + 
\frac{v_\perp^2}{2} + 
\phi(x,t) \Big]\,\dot t  \,,
\ee
It is straightforward to show from the Euler-Lagrange equations
\be
 \parfra{L'}{q'} - \ds{} \parfra{L'}{\dot q'} = 0
\ee
that $\theta$ is the fast gyro-angle, changing on the time scale $\eps$.

\subsubsection{Algebraic GY-transformations}

The second step of variational averaging requires a coordinate map $\tau^\eps: \Omega_\tn{gy} \to 
\Omega_\tn{I}'$, $q_\tn{gy} \mapsto q'$ which eliminates the fast variable $\alpha \mapsto \theta$ 
from the Lagrangian \eqref{L:X}, order by order in $\eps$. The second transformation is thus
assumed to be a finite power series in $\eps$, defined by
\be \label{transf2}
 q' = \tau^\eps(q_\tn{gy}) := q_\tn{gy} + \sum_{n=1}^{N+1} \eps^n\, \bfG_n(q_\tn{gy}) \,, 
\ee
where $N\geq 0$ denotes the order of the transformation and the $\bfG_n: \Omega_\tn{gy}\to 
\Omega_\tn{I}'$ are smooth maps, the so-called ``generating functions'' or 
generators of the transformation. They should be bounded uniformly in $\eps$, such that 
$\lim_{\eps\to0}\tau^\eps = \tau^0$ is the identity. Note that one needs $N+1$ generators in the 
$N$-th order transformation and that these generators occur merely as coefficients in the 
$\eps$-series (algebraic dependence on the generators). The GY-coordinates $q_\tn{gy} = 
(q_{\tn{gy},i})_{1\leq i \leq 7}$ and the generators $\bfG_n = (G_{n,i})_{1\leq i \leq 7}$ are 
denoted by
\begin{align*}
 &(q_{\tn{gy},i})_{1\leq i \leq 3} = r\,,\qquad\: q_{\tn{gy},4} = q_\para\,,\qquad\:\: 
q_{\tn{gy},5} = q_\perp\,,\qquad\: q_{\tn{gy},6} = \alpha\,,\qquad\:\:\: q_{\tn{gy},7} = 
t\,,
 \\[3mm]
 &(G_{n,i})_{1\leq i \leq 3} = \varrho_n\,\qquad G_{n,4} = G_n^\para\,,\qquad G_{n,5} = 
G_n^\perp\,,\qquad G_{n,6} = G_n^\theta\,,\qquad G_{n,7} = 0\,,
\end{align*}
where $r\in \RR^3$ denotes the GY-position, $q_\para$ and $q_\perp$ are the respective parallel and 
perpendicular GY-velocities and $\alpha$ stands for the gyro-angle. The time coordinate $t$ rests 
untransformed since we assume its generators to be zero at all orders. Component-wise, the 
transformation \eqref{transf2} thus reads
\be \label{X(Y)}
 \begin{aligned}
  x &= r + \sum_{n=1}^{N+1} \eps^n\, \varrho_n(q_\tn{gy})\,,
  \\
  v_\para &= q_\para + \sum_{n=1}^{N+1} \eps^n\, G_n^\para(q_\tn{gy})\,,
  \\
  v_\perp &= q_\perp + \sum_{n=1}^{N+1} \eps^n\, G_n^\perp(q_\tn{gy})\,,
  \\
  \theta &= \alpha + \sum_{n=1}^{N+1} \eps^n\, G_n^\theta(q_\tn{gy})\,,
  \\[1mm]
  t &= t\,.
 \end{aligned}
\ee
Moreover, from the definition of the tangent map one obtains
\be \label{dotx(Y)}
 \dot x = \dot r + \sum_{n=1}^{N+1} \eps^n\, \dot \varrho_n(q_\tn{gy}, \dot 
q_\tn{gy})\,,\qquad\quad 
\dot \varrho_n := \parfra{\varrho_n}{q_\tn{gy}} \cdot \dot q_\tn{gy}\,. 
\ee
Starting from \eqref{L:X} the tangent map leads to the extended Lagrangian $L^\eps$ in the 
variables~$q_\tn{gy}$,
\be \label{L:Y}
L^\eps(q_\tn{gy},\dot q_\tn{gy}) := L'(T \tau^\eps(q_\tn{gy},\dot q_\tn{gy})) = L'( 
\tau^\eps(q_\tn{gy}) , D \tau^\eps (q_\tn{gy})\cdot \dot q_\tn{gy}))\,.
\ee
If we assume sufficiently regular potentials $A$ and $\phi$, the definition of $\tau^\eps$ as 
a power series in \eqref{transf2} translates to a Taylor expansion of $L'$ around 
$(q_\tn{gy},\dot q_\tn{gy})$, 
leading to a representation of $L^\eps$ in the form \eqref{intro:Lgy}. The generators $\bfG_n$ are 
still undetermined in this formulation. As outlined by Kruskal and Littlejohn 
\cite{Kruskal,Little1983}, they can be chosen order by order such that the truncated Lagrangian 
$L_\tn{gy}^{(N)}$ is independent of the GY-angle $\alpha$.

\begin{remark}
The gyro-transformation (GT) that leads to the Lagrangian 
\eqref{intro:Lgy} will be composed of two transformations, $\tau_\tn{gy}^\eps = 
\tau' \circ \tau^\eps $, where $\tau'$ is the ``preliminary'' transformation independent of $\eps$ 
and $\tau^\eps$ denotes the algebraic GY-transformation \eqref{transf2}. 
Even though $\tau_\tn{gy}^\eps$ is a composition, it must not be confused with the ``two-step'' GT 
\cite{Brizard_rev2007}, where only the static 
$B_0$ is considered at first (guiding-center problem) and only after the dynamical fields 
$B_1$ and $E$ are taken into account. Indeed, the two-step GT is really a three-step GT since 
the preliminary transformation $\tau'$ is applied also in this case. Our procedure corresponds 
to what is known as the ``one-step'' GT.
\end{remark}

\section{Main results} \label{sec:results}

\subsection{Preliminaries} \label{subsec:prelim}

The main results have been arranged into three categories which are presented in the following 
three subsections: section \ref{subsec:prelim} contains a preliminary result 
on the existence of solutions to the initial value problem \eqref{ivp}, under the here used 
regularity assumptions on the electromagnetic potentials:

\begin{assumption} \label{assumps}
 For $N\geq 0$ we suppose ${A_0\in 
C^{N+3}(\Omega_x)}$, 
${A_1\in C^{N+2}(\Omega_x \times \RR)}$ for the vector potential and $\phi \in 
C^{N+1}(\RR;C^{N+2}(\Omega_x))$ for the electrostatic potential.
\end{assumption}

In section \ref{subsec:exist} the first main result Theorem \ref{thm0} shows that this 
regularity is sufficient for the existence of a truncated Lagrangian $L_\tn{gy}^{(N)}$ in 
\eqref{intro:Lgy}, independent of the gyro-angle, for arbitrary order $N$. Finally, section 
\ref{subsec:res2} concerns the 
error in the averaged dynamics due to truncation of the true Lagrangian. We give an exact 
definition of a gyrokinetic equation and compute a strong error bound for its solution 
with respect to the solution of the Vlasov equation \eqref{Vlasov} in our second main result, 
Theorem \ref{thm}. Let us start with 
some useful definitions:

\begin{definition} \label{def:Lgc}
 ({\bf Guiding-center Lagrangian}.)
 The guiding-center (GC) Lagrangian is defined as
 \be \label{def:Lgc}
  L_{\tn{gc}}(q_\tn{gy},\dot q_\tn{gy}) := \Big[ q_\para b_0(r) + \frac{A_0(r)}{\eps}  \Big] 
 \cdot \dot r + \eps\,\frac{q_\perp^2}{2|B_0(r)|}\,\dot \alpha  - \Big[ \frac{q_\para^2}{2} + 
 \frac{q_\perp^2}{2} \Big]\,\dot t \,.
 \ee
\end{definition}

\begin{definition} \label{def:equiv}
 ({\bf Equivalence of Lagrangians}.)
Two Lagrangians $L^*,L$ defined on $TM$ are equivalent, $L^*\sim L$, if there exists a 
function $S: M \to \RR$ such that ${L^* = L + \pa S/\pa q \cdot \dot q}$ in some coordinates $q$. 
Equivalent Lagrangians 
lead to the same Euler-Lagrange equations.
\end{definition}

\begin{definition} \label{def:gavg}
 ({\bf Gyro-average}.)
 The gyro-average and fluctuations 
of a function $G$ that is $2\pi$-periodic in $\alpha$ are defined by
\be \label{def:gyro}
 \gavg{G}(\cdot) := \frac{1}{2\pi} \int_0^{2\pi} 
G(\cdot,\alpha)\,d\alpha\,,\qquad\quad 
\widetilde G := G - 
\gavg{G}\,.
\ee
\end{definition}

\begin{assumption} \label{assumps:2}
 In the initial value problem \eqref{ivp} we denote by $\rho$ the radius of the largest ball 
in $\Omega_x$ containing $x_0$, that is $\rho := 
\sup_{R\in\RR}(\{x\in\RR^3:|x-x_0|<R\,,\: x\in\Omega_x\})$. 
Moreover, 
the velocity space is bounded by a maximal kinetic energy, ${\Omega_v =\{v\in\RR^3:|v|^2 
< v_\mathrm{max}^2\}}$, and ${\rho_\mathrm{kin} := (v_\mathrm{max}^2 - |v_0|^2)/2}$.
\end{assumption}

\begin{lemma} \label{lem:1}
Let $\eps >0$. Under the assumptions \ref{assumps} and \ref{assumps:2} the initial value 
problem \eqref{ivp} has a unique solution which exists for $t \in [t_0,t_0 + T] $ with
$T=\min(\rho/v_\mathrm{max}\,,\,\rho_\mathrm{kin}/(E_\mathrm{max}v_\mathrm{max}))$,
independent of $\eps$. 
\begin{proof}
 Due to assumption \ref{assumps} the fields in \eqref{ivp} are continuous on a bounded domain 
$\Omega$; hence the electric field has an upper bound, $|E|<E_\mathrm{max}$. We only need to check 
at which time the solution will leave $\Omega$. Integrating the first equation and 
taking the norm yields
$$
 |x(t) - x_0| \leq (t-t_0)v_\mathrm{max} < \rho \qquad\implies\qquad (t-t_0) < 
\frac{\rho}{v_\mathrm{max}}\,.
$$
Multiplying the second equation by $v$ and integrating in time leads to
$$
 \frac{1}{2}(|v(t)|^2 - |v_0|^2) \leq (t-t_0)E_\mathrm{max} v_\mathrm{max} < \rho_\mathrm{kin} 
\qquad\implies\qquad (t-t_0) < 
\frac{\rho_\mathrm{kin}}{E_\mathrm{max}v_\mathrm{max}}\,,
$$
which determines the time interval $T$. 
\end{proof}
\end{lemma}

\subsection{Existence of an algebraic GY-map $\tau^\eps$} \label{subsec:exist}

\begin{prop} \label{prop:1} ({\bf Series-expansion of $L^\eps$}.)
 Under assumption~\ref{assumps} the Lagrangian $L^\eps$ in 
\eqref{L:Y} is equivalent to the series expansion
 \be \label{prop:1:result}
  L^\eps \sim \frac{1}{\eps} L_{-1} + L_0 + \sum_{n=1}^{N} \eps^n L_n + O(\eps^{N+1}) \,,
 \ee
 with the terms
\begin{align*}
 L_{-1} &= A_0 \cdot \dot r\,,
 \\[1mm]
 L_0 &= (q_\para b_0 + q_\perp c_0 - \varrho_1 \times B_0 + A_1 ) \cdot \dot r - \Big(
\frac{q_\para^2}{2} + \frac{q_\perp^2}{2} + \phi \Big)\,\dot t\,,
 \\[2mm]
 L_{1\leq n\leq N} &= \Big[ G^\para_{n} b_0 + G^\perp_{n}\, c_0 - \varrho_{n+1} \times 
B_0 - \varrho_n \times B_1 + \cQ_n \Big] \cdot \dot r
 \\[2mm]
 &\quad  - ( q_\para G^\para_{n} + q_\perp 
G^\perp_{n} - \varrho_n \cdot E )\,\dot t + \cL_n \,.
\end{align*}
Here, the linear form $\cQ_n$ and the Lagrangian $\cL_n$ are given in 
\eqref{def:cQgy} and \eqref{def:cLgy}, respectively. For $n=1$ and $n=2$ they can be
written in terms of the fields $B_0$, $B_1$ and $E$ (gauge-invariance).
\begin{proof}
 The proof is written in section \ref{proof:prop1}.
\end{proof}
\end{prop}

\begin{theorem} ({\bf Existence of $\tau^\eps$.}) \label{thm0}
 Under assumption~\ref{assumps}, for all $N\geq 1$, there exist generators $\bfG_n \in 
C^2(\Omega_\tn{gy})$, $1\leq n\leq N+1$, of the algebraic GY-transformation $\tau^\eps$ such that 
$L^\eps$ from \eqref{prop:1:result} is equivalent to $L^\eps \sim L_\tn{gy}^{(N)} + O(\eps^{N+1})$, 
where the GY-Lagrangian reads
\be \label{thm0:result}
 L_\tn{gy}^{(N)} = L_\tn{gc} + A_1 \cdot \dot r - \phi\,\dot t + \eps^2\,\delta\mu^{(N)}\,\dot 
\alpha\,.
\ee
Here, $L_\tn{gc}$ denotes the guiding-center Lagrangian defined in \eqref{def:Lgc} and $\delta 
\mu^{(N)}:\Omega_\tn{gy} \to \RR$ is the $N$-th order correction to the magnetic moment $\mu = 
q_\perp^2/(2|B_0|)$, independent of $\alpha$. 
\begin{proof}
 The proof is written in section \ref{proof:thm0}.
\end{proof}
\end{theorem}

\begin{remark}
 The existence result from Theorem \ref{thm0} does not imply that the transformation $\tau^\eps$ 
exists as $N\to \infty$, because we cannot say that the series \eqref{transf2} converges in this 
limit. Convergence of the series would mean that a gyro-center of the motion exists globally. This 
is true for the constant field case $B=const.$ and $E=const.$ but it is not clear in the 
non-homogeneous case. In practice, however, only low orders $N\leq2$ are important for numerical 
purposes.
\end{remark}

\subsection{An error estimate for gyrokinetics}  \label{subsec:res2}

Due to the Euler-Lagrange equation \eqref{conserve}, the result in Theorem \ref{thm0} leads to the 
conservation of the generalized magnetic moment $\wh\mu$ during the GY-motion, where
\be \label{def:mu}
\wh \mu := \frac{q_\perp^2}{2|B_0|} + \eps\,\delta \mu^{(N)}\,.
\ee
In order to reduce the dimension of the problem, $\wh\mu$ must be adopted as 
one of the coordinates. In particular, we shall assume that there is a one-to-one correspondence 
$\wh\tau: \wh\mu \mapsto q_\perp$, which is the case in all of the results presented in 
section \ref{sec:express}. The full GY-transformation from $x$-$v$-$t$-coordinates (extended phase 
space $\Omega_\tn{I}$) to the GY-coordinates $\wh q_\tn{gy} \in 
\wh \Omega_\tn{gy}$ with generalized magnetic moment, hence $\wh 
q_\tn{gy} = (r,q_\para,\wh\mu,\alpha,t)$, is given by the composition
\be \label{full:transf}
 \tau_\tn{gy}^\eps: \wh \Omega_\tn{gy} \to \Omega_\tn{I}\,,\qquad\quad \tau_\tn{gy}^\eps = \tau' 
\circ \tau^\eps \circ \wh\tau\,.
\ee
It follows from Theorem \ref{thm0} that the exact dynamics can be obtained from the Lagrangian
\be \label{Leps:res2}
 L^\eps \sim L_\tn{gy}^{(N)} + O(\eps^{N+1})\,,
\ee
which is now written in terms of the coordinates $\wh q_\tn{gy}$ as
\be \label{L:trunc}
 L_\tn{gy}^{(N)} = \frac{1}{\eps}\, A^* \cdot \dot r - H_\tn{gy}\,\dot t + \eps\,\wh \mu\,\dot 
\alpha\,, 
\ee
with the auxiliary potential $A^*$ and the GY-Hamiltonian $H_\tn{gy}$ defined by
\begin{align*}
 A^* &:= A_0 + \eps\,A_1 + \eps\, q_\para \, b_0\,,
 \\[1mm]
 H_\tn{gy} &:= \frac{q_\para^2}{2} + \wh \mu\,|B_0| + \phi + \eps\,\delta H^{(N)} \,.
\end{align*}
Here, the Hamiltonian correction $\delta H^{(N)}$ stems from the transformation of the term 
$q_\perp^2/2$ under the map $\wh\tau$. It is a remarkable feature that $\delta H^{(N)}$ is the only 
term in the Lagrangian that changes with the order $N$ of the transformation. This means in 
particular that, in the coordinates $\wh q_\tn{gy}$, the non-time components 
$(A^*/\eps,0,0,\eps\,\wh\mu)$ of the GY-symplectic form are the same for all $N$.

Setting $\dot t = 1$ in \eqref{L:trunc}, a straightforward computation yields the Euler-Lagrange 
equations of 
\eqref{Leps:res2}, here stated with initial conditions, and with the abbreviations $B^* = \nabla 
\times A^*$, $B^*_\para = B^* \cdot b_0$ and $E^* = E - \wh \mu \,\nabla|B_0| - \eps\,\nabla \delta 
H^{(N)}$,
 \be \label{dt:true}
 (P^\eps)
\left\{
\begin{aligned}
  \dt r &= \frac{1}{B^*_\para}\Big( q_\para + \eps\, \parfra{\delta H^{(N)}}{q_\para} \Big) B^* + 
\eps\frac{1}{B^*_\para} E^* 
\times b_0 + O(\eps^{N+2})\,, &&\qquad 
r(t_0) = r_0\,,
 \\[0mm]
 \dt {q_\para} &= \frac{1}{B^*_\para}\, B^* \cdot E^* + 
O(\eps^{N+1})\,, &&\qquad q_\para(t_0) 
= q_{\para0}
 \\[0mm]
 \dt{ \wh \mu} &= O(\eps^N) \,, &&\qquad \wh\mu(t_0) = \wh\mu_0\,,
 \\[1mm]
 \dt \alpha &= \frac{|B_0|}{\eps} + \parfra{\delta H^{(N)}}{\wh\mu} + O(\eps^N)\,, &&\qquad 
\alpha(t_0) = 
\alpha_0\,,
\end{aligned}
\right.
\ee
Let $f_\tn{gy}:\Omega_\tn{gy} \to \RR_+$ denote the unique 
function which is constant along the solutions of $(P^\eps)$, with initial 
condition $f_\tn{gy}(t=t_0) = f_{0,\tn{gy}}$ strictly positive. Since $(P^\eps)$ is merely a 
reformulation of the 
initial-value problem \eqref{ivp} in the coordinates $\wh q_\tn{gy}$ via the map 
$\tau_\tn{gy}^\eps$, we have $f_\tn{gy} = f\circ \tau_\tn{gy}^\eps$, where $f$ is the unique 
solution of 
the Vlasov equation \eqref{Vlasov} with initial condition $f_0 = 
f_{0,\tn{gy}} \circ (\tau_\tn{gy}^\eps)^{-1}$. Regarding existence of $f_\tn{gy}$ we remark:

\begin{lemma} \label{lem:3}
 A solution of the problem $(P^\eps)$ exists, is unique and continuous on the interval 
$[t_0,t_0+T]$, where $T$ is given in Lemma \ref{lem:1}. 
\begin{proof} This follows from Lemma \ref{lem:1} and the fact that $(P^\eps)$ is 
equivalent to the initial value problem \eqref{ivp} if ${\wh q_\tn{gy,0} = 
(\tau_\tn{gy}^\eps)^{-1}(x_0,v_0,t_0)}$. The inverse of the algebraic GY-map 
${\tau_\tn{gy}^\eps = \tau' \circ \tau^\eps \circ \wh \tau}$ exists at least locally due to the 
fact $\tau^\eps$ is continuously differentiable because all generators
are (Theorem \ref{thm0}).
\end{proof}
\end{lemma}

If one truncates the residual terms of order $O(\eps^{k\geq N})$ in \eqref{dt:true}, one obtains 
the ``decoupled dynamics'', 
which are the Euler-Lagrange equations of the truncated Lagrangian \eqref{L:trunc}. A decoupled 
solution is denoted with an overbar, hence the decoupled problem is written as
 \be \label{dt:avg}
 (\overline P^\eps)
\left\{
\begin{aligned}
  \dt{\overline r} &= \frac{1}{B^*_\para}\Big( \overline q_\para  + \eps\, \parfra{\delta 
H^{(N)}}{q_\para} 
\Big) B^* + \eps\frac{1}{B^*_\para} E^* 
\times b_0 \,, &&\qquad 
 \overline r(t_0) = \overline r_0\,,
 \\[0mm]
 \dt{\overline q_\para } &= \frac{1}{B^*_\para}\, B^* \cdot E^* \,, &&\qquad 
\overline q_\para (t_0) = \overline q_{\para0}
 \\[0mm]
 \dt{\overline{\wh \mu}} &= 0 \,, &&\qquad \overline{\wh\mu}(t_0) = \overline{\wh\mu}_0\,,
 \\[1mm]
 \dt{\overline{\alpha}} &= \frac{|B_0|}{\eps} + \parfra{\delta H^{(N)}}{\wh\mu} \,, &&\qquad 
 \overline \alpha(t_0) = \overline \alpha_0\,,
\end{aligned}
\right.
\ee

\begin{definition} ({\bf Gyrokinetic equation}.)
 A solution $F: \wh \Omega_\tn{gy} \to \RR_+$ of a gyrokinetic equation is a strictly positive 
function, 
constant along the solutions of the decoupled dynamics~\eqref{dt:avg}, 
\be \label{def:GK}
\left\{
\begin{aligned}
 &\parfra{F}{t} + \dt{\overline r} \cdot \parfra{F}{r} + \dt{\overline q_\para 
}\,\parfra{F}{q_\para} + \dt{\overline{\alpha}}\,\parfra{F}{\alpha} = 0\,,
 \\[1mm]
 &F(t=t_0) = F_0\,.
 \end{aligned}
 \right.
\ee
\end{definition}

\begin{lemma}
 The gyro-average $\gavg F$ and fluctuations $\widetilde F$ of a solution to \eqref{def:GK} 
evolve independently in time (decoupling); they satisfy
\begin{align}
 &\parfra{\gavg F}{t} + \dt{\overline r} \cdot \parfra{\gavg F}{r} + \dt{\overline q_\para 
}\,\parfra{\gavg F}{q_\para} = 0\,, && \gavg F(t=t_0) = \gavg{ 
F_0}\,,  \label{GK:avg}
 \\[2mm]
 &\parfra{\widetilde F}{t} + \dt{\overline r} \cdot \parfra{\widetilde F}{r} + \dt{\overline 
q_\para 
}\,\parfra{\widetilde F}{q_\para} + \dt{\overline{\alpha}}\,\parfra{\widetilde F}{\alpha} = 0\,, 
 && \widetilde F(t=t_0) = \widetilde F_0\,.
\end{align}
\begin{proof}
 This is an immediate consequence of the definition \ref{def:gavg} of the gyro-average and the fact 
that the direction field (right-hand-side) in \eqref{dt:avg} is independent of the gyro-angle 
$\overline \alpha$.
\end{proof}
\end{lemma}

In what follows we denote by $\bfz := (r,q_\para,\wh\mu)$ the slow components of the phase space 
variables $\wh q_\tn{gy}$. In the decoupled dynamics the ``slow trajectories'' $\overline \bfz(t) 
:= (\overline 
r(t),\overline q_\para(t), \overline{\wh\mu}(t))$ evolve independently from the gyro-angle 
$\overline \alpha(t)$, which varies with a frequency 
$1/\eps$. From \eqref{dt:true} and \eqref{dt:avg} we can extract two 
subproblems for the slow variables,
\be \label{Pprobs:sub}
(P_\bfz^\eps)\left\{
\begin{aligned}
 &\dt{\bfz} = \Lambda(\bfz,t,\eps) + \eps^N\, 
S(\bfz, \alpha,t,\eps)\,,
 \\[1mm]
 &\bfz(t_0) = \bfz_0\,,
\end{aligned}
\right.
\qquad\quad
(\overline P_\bfz^\eps)\left\{
\begin{aligned}
 &\dt{\overline\bfz} = \Lambda(\overline\bfz,t,\eps) \,,
 \\[1mm]
 &\overline\bfz(t_0) = \overline\bfz_0\,.
\end{aligned}
\right.
\ee
Here, $\Lambda$ stands for the direction field for the slow variables in \eqref{dt:avg} and $S$ 
comprises 
the residual terms $O(\eps^{k\geq N})$ of \eqref{dt:true} for the slow variables.

\begin{remark}
 The direction field $\Lambda$ is independent of $\alpha$. From assumption \ref{assumps} we 
 deduce that it is a $C^1$-function of $(\bfz,t,\eps)$. In particular, $\Lambda$ is 
 Lipschitz in $\bfz$, uniformly in $(t,\eps)$,
 $$
 ||\Lambda(\bfy,t,\eps) - \Lambda(\bfz,t,\eps)|| \leq \ell_\Lambda\, || \bfy 
-  \bfz ||\,,
 $$
 for some vector norm $||\cdot ||$, where the Lipschitz constant $\ell_\Lambda$ is independent of 
$(t,\eps)$.
\end{remark}

\begin{remark}
 The residual term $S$ depends on $\alpha$. It is obtained from the 
$O(\eps^{N+1})$-terms in the Lagrangian \eqref{Leps:res2} via the Euler-Lagrange equations. Since 
these $O(\eps^{N+1})$-terms are the residuals in the Taylor expansion \eqref{Taylor}, they 
are $C^1$-functions of $(\bfz,\alpha,t)$. Therefore $S$, being a power series in $\eps$, is 
continuous in $(\bfz,\alpha,t,\eps)$.
\end{remark}

\begin{lemma} \label{lem:2}
 Consider the problems $( P_\bfz^\eps)$ and $(\overline P_\bfz^\eps)$ for the slow 
$GY$-variables on the interval $t\in I=[t_0,t_1]$ with $t_1\leq T$ (Lemma \ref{lem:3}), then
$$
 || \overline \bfz(t) - \bfz(t) ||  \leq  \eps^N\, 
\frac{||S||_{\infty,\eps}}{\ell_\Lambda} \big(e^{ 
\ell_\Lambda(t-t_0)} - 1\big) + || \overline \bfz_0 - \bfz_0 ||\,  e^{ 
\ell_\Lambda(t-t_0)} \,,
$$
where $||S||_{\infty,\eps} = \max_{\wh \Omega_\tn{gy} \times (0,\eps_\tn{max}]} ||S|| = 
O(1)$ as $\eps \to 0$.
\begin{proof}
 The proof is written in section \ref{proof:lem:2}.
\end{proof}
\end{lemma}

\begin{theorem} \label{thm}
Let $\gavg F$ denote the unique solution of the averaged part 
\eqref{GK:avg} of the gyrokinetic equation on the interval $I=[t_0,t_1]$, with initial condition 
$\gavg{F_0}$. Moreover, let ${f_\tn{gy} = f \circ \tau_\tn{gy}^\eps}$, where $f$ is the unique 
solution of the Vlasov equation \eqref{Vlasov} with initial data $f_0$, hence ${f_{0,\tn{gy}} 
= f_0 \circ \tau_\tn{gy}^\eps}$, where $\tau_\tn{gy}^\eps$ denotes the transformation 
\eqref{full:transf} of order $N$. Suppose 
\begin{enumerate}
 \item $\gavg{F_0} = \gavg{f_{0,\tn{gy}}}$, \hspace{4mm}  Lipschitz with constant $\ell_0$,
 \item $f_{0,\tn{gy}} = \gavg{f_{0,\tn{gy}}} + \eps^N \widetilde f_{0,\tn{gy}}$, \hspace{4mm} with 
$\widetilde f_{0,\tn{gy}} = O(1)$ as $\eps \to 0$ continuous\,.
\end{enumerate}
Then, denoting $\bfz = (r,q_\para,\wh\mu)$, for $t \in I$ one has
$$
 \max_{\bfz,\alpha} \big| \gavg F(\bfz,t) - f_\tn{gy}(\bfz,\alpha,t) \big| \leq \eps^N\, 
C(t) \,,
$$
with 
$$
 C(t) = \ell_0\,\frac{||S||_{\infty,\eps}}{\ell_\Lambda} \big( e^{\ell_\Lambda(t-t_0)} - 1 \big) + 
\max_{\bfz,\alpha}  \big| \widetilde f_{0,\tn{gy}}(\bfz,\alpha) \big|\,,
$$
where the function $S$ is the one from Lemma \ref{lem:2}.
\begin{proof}
 Let $\Phi_{s,t}: \wh\Omega_\tn{gy} \to \wh\Omega_\tn{gy}$ stand for the local flow 
map\footnote{We have $\Phi_{t,s} = \Phi_{s,t}^{-1}$, since $\tn{id}_{\wh \Omega_\tn{gy}} = 
\Phi_{t,t} = 
\Phi_{t,s} \circ \Phi_{s,t}$ by the semi-group property of the local flow.} of problem $(P^\eps)$, 
i.e $\Phi_{s,t}(\bfz,\alpha)$ is the solution of $(P^\eps)$ at time $s$ which is at $(\bfz,\alpha)$ 
at time $t$, and $\Phi_{t,t} = \tn{id}_{\wh\Omega_\tn{gy}}$. We shall 
denote the ``slow'' components of the flow by $\bfZ_{s,t}$, corresponding to $\bfz$ for 
the slow variables, i.e. $\bfZ_{t,t}(\bfz,\alpha) = \bfz$ and $\bfZ_{t,t_0}(\bfz_0,\alpha_0) = 
\bfz(t)$, solution of the subproblem $(P^\eps_\bfz)$ written in \eqref{Pprobs:sub}.
Using that $f_\tn{gy}$ is constant along solutions of $(P^\eps)$ we may write
\begin{align*}
 f_\tn{gy}(\bfz,\alpha,t) &= f_\tn{gy}(\Phi_{t,t}(\bfz,\alpha),t)
 \\[3mm]
 &= f_\tn{gy}(\Phi_{t_0,t}(\bfz,\alpha),t_0)
 \\[3mm]
 &= f_{0,\tn{gy}}(\Phi_{t_0,t}(\bfz,\alpha))
 \\[2mm]
 &=  \gavg{f_{0,\tn{gy}}}(\bfZ_{t_0,t}(\bfz,\alpha)) + \eps^N\,
\widetilde f_{0,\tn{gy}}(\Phi_{t_0,t}(\bfz,\alpha)) \,.
\end{align*}
Our aim is to compare this expression to $\gavg F$, solution of \eqref{GK:avg}. For this let us 
denote by $\overline \bfZ_{s,t}$ the flow map of the subproblem $(\overline P^\eps_\bfz)$ from 
\eqref{Pprobs:sub}, i.e. $\overline \bfZ_{t,t}(\overline \bfz) = \overline \bfz$ and 
$\overline \bfZ_{t,t_0}(\overline \bfz_0) = \overline \bfz(t)$. Since $\gavg F$ is constant along 
solutions of $(\overline P^\eps_\bfz)$ we have
\begin{align*}
 \gavg F (\bfz,t) &= \gavg F(\overline \bfZ_{t,t}(\bfz),t)
 \\[2mm]
 &= \gavg F(\overline \bfZ_{t_0,t}(\bfz),t_0)
 \\[2mm]
 &= \gavg{F_0}(\overline \bfZ_{t_0,t}( \bfz))
 \\[2mm]
 &= \gavg{f_{0,\tn{gy}}}(\overline \bfZ_{t_0,t}(\bfz))\,.
\end{align*}
Therefore, using the Lipschitz continuity of $\gavg{f_{0,\tn{gy}}}$ we obtain
\begin{align*}
 \big| \gavg F (\bfz,t) - f_\tn{gy}(\bfz,\alpha,t) \big| &= \big| \gavg{f_{0,\tn{gy}}}(\overline 
\bfZ_{t_0,t}(\bfz)) -  \gavg{f_{0,\tn{gy}}}(\bfZ_{t_0,t}(\bfz,\alpha)) - \eps^N\,
 \widetilde f_{0,\tn{gy}}(\Phi_{t_0,t}(\bfz,\alpha)) \big|
 \\[3mm]
 &\leq \big| \gavg{f_{0,\tn{gy}}}(\overline 
\bfZ_{t_0,t}(\bfz)) -  \gavg{f_{0,\tn{gy}}}(\bfZ_{t_0,t}(\bfz,\alpha))  \big| + \eps^N 
\big| \widetilde f_{0,\tn{gy}}(\Phi_{t_0,t}(\bfz,\alpha)) \big|
 \\[3mm]
 &\leq \ell_0\,|| \overline \bfZ_{t_0,t}(\bfz)) - \bfZ_{t_0,t}(\bfz,\alpha) || + \eps^N 
\big| \widetilde f_{0,\tn{gy}}(\Phi_{t_0,t}(\bfz,\alpha)) \big|\,.
\end{align*}
The continuity of $\widetilde f_{0,\tn{gy}}$ leads to a bound for the second term.
The difference in the flow functions can be estimated from Lemma \ref{lem:2}, 
 $$
 \forall\,(\bfz,\alpha,t) \in \Omega_\tn{gy} \cap I :\qquad || \overline 
\bfZ_{t_0,t}(\bfz)) - \bfZ_{t_0,t}(\bfz,\alpha) || \leq  \eps^N\, 
\frac{||S||_{\infty,\eps}}{\ell_\Lambda} 
\big(e^{ \ell_\Lambda(t-t_0)} - 
1\big) \,.
 $$
\end{proof}
\end{theorem}

\section{Expressions for $\delta \mu^{(N)}$, $\delta H^{(N)}$ and the generators}  
\label{sec:express}

Theorem \ref{thm0} states the existence of algebraic GY-maps $\tau^\eps$ that lead to the reduced 
dynamics implied by the Lagrangian \eqref{thm0:result}. Here we give some concrete examples of such 
transformations for the two different scalings \eqref{scalings} of $\eps_B$ and for the orders 
$N=1$ and $N=2$, 
respectively. We stress that the choice for the generators in these transformations is not 
unique for two reasons:
\begin{enumerate}
 \item the equivalence of Lagrangians that differ by a ``total time derivative'' allows us to add 
or subtract such a term,
 \item gyro-averages of the generators could be kept in the Lagrangian rather than in the 
transformation; one then still obtains a set of decoupled equations of motion, albeit a rather 
different one, c.f. the discussion in remark \ref{rem:5}.
\end{enumerate}

Moreover, we state the derived expressions for the correction $\delta \mu^{(N)}$ to the magnetic 
moment, the map $\wh\tau$ between $\wh \mu$ and $q_\perp$, as well as the correction $\delta 
H^{(N)}$ to the Hamiltonian. The proof of the following Lemmas is left as an exercise to the 
reader; it consists of performing the steps described in the proof of Theorem \ref{thm0}.

\begin{lemma} \label{lemma:4} ({\bf Small background variations} $\eps_B = \eps$, $N=1$.)
 In this case possible generators read
 \be \label{lemma:4:eqs}
\begin{aligned}
 \varrho_1 &= \frac{q_\perp}{|B_0|}\,a_0\,,    
 \\[1mm]
 G^\para_1 &= - \frac{q_\perp}{|B_0|} B_1 \cdot c_0 \,,   
 \\[1mm]
 G^\perp_1 &= \frac{q_\para}{|B_0|} B_1 \cdot c_0 + \frac{1}{|B_0|} E \cdot a_0  \,, 
 \\[1mm]
 \varrho_2 &= \Big[ \frac{q_\para}{|B_0|^2} B_1 \cdot c_0 - \frac{q_\perp}{|B_0|^2} B_1 \cdot b_0 + 
\frac{1}{|B_0|^2} E \cdot a_0 \Big] a_0  + (\varrho_2 \cdot b_0)\,b_0 + 
\frac{q_\perp}{|B_0|} G_1^\theta\,c_0\,,  
\end{aligned}
\ee
where $\varrho_2 \cdot b_0$ and $G_1^\theta$ 
are arbitrary. This leads to $\delta \mu^{(1)} = 0$, therefore $\wh \mu 
= \mu$, the map $\wh \tau:\wh \mu\mapsto q_\perp$ is given by $q_\perp = \sqrt{2\,\wh \mu 
\,|B_0|}$ and the Hamiltonian correction is zero, 
$$
\delta H^{(1)} = 0\,.
$$
\end{lemma}

\begin{lemma} \label{lemma:4b}({\bf Small background variations} $\eps_B = \eps$, $N=2$.)
In this case  possible generators are the functions in \eqref{lemma:4:eqs} along with
\be
\begin{aligned}
 \varrho_2 \cdot b_0 &= \parfra{S_2^{**}}{q_\para} \,, 
 \\[0mm]
  G^\theta_1 &= \frac{|B_0|}{q_\perp}\, \parfra{S_2^{**}}{q_\perp} \,,  
 \\[2mm]
 G^\para_2 &= (\varrho_2 \times B_1) \cdot b_0 - \cQ_2^{**} \cdot b_0 \,,
 \\[2mm]
 G^\perp_2 &= - \frac{q_\para}{q_\perp} G^\para_2 + \frac{1}{q_\perp} \varrho_2 \cdot E + 
\frac{q_\perp}{2 |B_0|^2} 
a_0 \cdot \nabla E \cdot a_0 + \frac{1}{q_\perp} \parfra{S_2^{**}}{t}\,, 
 \\[1mm]
 \varrho_3 &= \frac{1}{|B_0|} \Big[ G^\perp_2 -  (\varrho_2 \times B_1) \cdot c_0 + 
\cQ_2^{**} \cdot c_0 \Big]\,a_0 
 \\[1mm]
 &\quad + (\varrho_3 \cdot b_0)\,b_0 + \frac{1}{|B_0|} \Big[(\varrho_2 \times B_1) \cdot a_0 - 
\cQ_2^{**} \cdot 
a_0 \Big]\, c_0\,,
\end{aligned}
\ee
where $\cQ_2^{**}$ is given by
\be \label{Pgi2star:4}
 \begin{aligned}
 \cQ_2^{**} &= \frac{q_\perp^2}{2|B_0|^2} (a_0 \cdot \nabla B_0) \times a_0 
+ \frac{q_\perp^2}{2|B_0|^2}
(a_0 \cdot \nabla B_1) \times a_0 
 \\[1mm]
 &\quad - \frac{q_\para q_\perp}{|B_0|} \,a_0
\times (\nabla \times b_0) - \frac{q_\perp^2}{|B_0|}\,a_0   \times (\nabla \times c_0) - 
G^\perp_1\,G^\theta_1\,a_0 - q_\perp\,G^\theta_2\,a_0
 \\[1mm]
 &\quad - \frac{q_\perp}{2}\,(G^\theta_1)^2\,c_0 -\frac{q_\perp^2}{2|B_0|}\,\tn R \,.
\end{aligned}
\ee
with $\tn R = \nabla a_0 \cdot c_0 = \nabla e_2 \cdot e_1$ denoting the gyro-gauge term, 
$\varrho_3 \cdot b_0$ and $G_2^\theta$ are arbitrary and $S_2^{**}$ reads
$$
\begin{aligned}
 S_2^{**} &= - \frac{q_\para q_\perp}{|B_0|^2} B_1 \cdot a_0 + \frac{q_\perp}{|B_0|^2} E \cdot c_0 
\,.
\end{aligned}
$$
This leads to
 $$
 \delta \mu^{(2)} = \frac{q_\perp^2}{2|B_0|}\,\underbrace{\frac{(- B_1 \cdot b_0) 
}{|B_0|}}_{=:\sigma^{**}}\,,
 $$
where $\sigma^{**}$ is independent of $q_\perp$. Therefore, $\wh \mu = \mu(1+\eps\,\sigma^{**})$, 
the map $\wh \tau:\wh \mu\mapsto q_\perp$ is given by $q_\perp = \sqrt{2\,\wh \mu
\,|B_0|/(1+\eps\,\sigma^{**})}$ and the Hamiltonian correction reads 
$$
\delta H^{(2)} = -\wh \mu\,|B_0|\,\frac{\sigma^{**}}{1 + \eps\,\sigma^{**}}\,.
$$
\end{lemma}

\begin{remark} \label{rem:5}
 Standard second-order gyrokinetic Lagrangians in the long-wavelength approximation 
\cite{Tronko2017}, used for example in the codes GENE and ORB5 \cite{GENE,ORB5}, can be recovered 
from Lemma \ref{lemma:4b}. However, our choice of the generators differs from the 
conventional ones and leads to simpler equations of motion. For example, the polarization term 
$|\nabla \phi|^2$ usually appearing in gyrokinetic Hamiltonian functions at second order has been 
included in the generator $G_2^\perp$ in our formalism; it is hidden in the term $\varrho_2 \cdot 
E$ and does not play a role in the particle dynamics, which are derived from the Lagrangian 
\eqref{thm0:result} and are thus simpler. The polarization term re-appears only through the use of 
the GY-transformation, respectively its inverse, in the transformation to $x,v$-phase-space. This 
reflects our general strategy of keeping the particle dynamics as simple as possible by keeping a 
maximum number of terms in the generating functions, instead of the Lagrangian (see the proof 
section for more details). A new class of gyrokinetic numerical schemes based on this strategy 
could be envisioned.
\end{remark}

\begin{lemma} \label{lemma:2} ($\eps_B = 1$, $N=1$.)
 In this case possible generators read
\begin{align}
 \varrho_1 &= \frac{q_\perp}{|B_0|}\,a_0\,,    \nonumber
 \\[1mm]
 G^\para_1 &= - \frac{q_\perp}{|B_0|} B_1 \cdot c_0 + \frac{q_\perp^2}{2|B_0|} a_0 
\cdot \nabla b_0 \cdot c_0 - \frac{q_\para q_\perp}{|B_0|} ( \nabla \times b_0) 
\cdot c_0 - \frac{q_\perp^2}{2|B_0|}\, \tn R \cdot b_0 \,,   \nonumber
 \\[1mm]
 G^\perp_1 &= - \frac{q_\para}{q_\perp} G^\para_1 + \frac{1}{|B_0|}a_0 \cdot E  \,,  
\label{lemma:2:eq2}
 \\[1mm]
 \varrho_2 &= \Big[ \frac{G_1^\perp}{|B_0|} - \frac{q_\perp}{|B_0|^2} B_1 \cdot b_0 - 
\frac{q_\perp^2}{2|B_0|^3}\, 
a_0 \cdot \nabla |B_0| - \frac{q_\para q_\perp}{|B_0|^2} ( \nabla \times b_0) \cdot b_0 + 
\frac{q_\perp^2}{2|B_0|^2} \, \tn R \cdot c_0 \Big] a_0  
\nonumber   \\[1mm]
 &\quad + (\varrho_2 \cdot b_0)\,b_0 + \Big[ \frac{q_\perp\, G_1^\theta}{|B_0|} + 
\frac{q_\perp^2}{2|B_0|^2}\, \tn R \cdot a_0 \Big]\,c_0\,,  \nonumber
\end{align}
where $\tn R = \nabla a_0 \cdot c_0 = \nabla e_2 \cdot e_1$ is the gyro-gauge term and $\varrho_2 
\cdot b_0$ and $G_1^\theta$ are arbitrary. This leads to $\delta \mu^{(1)} = 0$, therefore $\wh \mu 
= \mu$, the map $\wh \tau:\wh \mu\mapsto q_\perp$ is given by $q_\perp = \sqrt{2\,\wh \mu 
\,|B_0|}$ and the Hamiltonian correction is zero, 
$$
\delta H^{(1)} = 0\,.
$$
\end{lemma}

\begin{lemma} \label{lemma:2b} ($\eps_B = 1$, $N=2$.)
 In this case possible generators are the functions in \eqref{lemma:2:eq2} along with
\be
\begin{aligned}
 \varrho_2 \cdot b_0 &= \parfra{S_2^*}{q_\para} - \frac{q_\perp^2}{2|B_0|^2}\, a_0 \cdot \nabla b_0 
\cdot a_0  \,, 
 \\[0mm]
  G^\theta_1 &= \frac{|B_0|}{q_\perp}\, \parfra{S_2^*}{q_\perp} + \frac{q_\perp}{2|B_0|}\, a_0 
\cdot 
\tn R \,,  
 \\[2mm]
 G^\para_2 &= (\varrho_2 \times B_1) \cdot b_0 - \cQ_2^* \cdot b_0 \,,
 \\[2mm]
 G^\perp_2 &= - \frac{q_\para}{q_\perp} G^\para_2 + \frac{1}{q_\perp} \varrho_2 \cdot E + 
\frac{q_\perp}{2 |B_0|^2} 
a_0 \cdot \nabla E \cdot a_0 + \frac{1}{q_\perp} \parfra{S_2^*}{t}\,, 
 \\[1mm]
 \varrho_3 &= \frac{1}{|B_0|} \Big[ G^\perp_2 -  (\varrho_2 \times B_1) \cdot c_0 + 
\cQ_2^* \cdot c_0 \Big]\,a_0 
 \\[1mm]
 &\quad + (\varrho_3 \cdot b_0)\,b_0 + \frac{1}{|B_0|} \Big[(\varrho_2 \times B_1) \cdot a_0 - 
\cQ_2^* \cdot 
a_0 \Big]\, c_0\,,
\end{aligned}
\ee
where $\varrho_3 \cdot b_0$ and $G_2^\theta$ 
are arbitrary and $S_2^*$ reads
$$
\begin{aligned}
 S_2^* &= - \frac{q_\para q_\perp}{|B_0|^2} B_1 \cdot a_0 + \frac{q_\perp}{|B_0|^2} E \cdot c_0 + 
\frac{q_\para 
q_\perp^2}{4|B_0|^2} ( a_0 \cdot \nabla b_0 \cdot a_0 ) 
 \\[0mm]
 &\quad - \frac{q_\para^2w_\perp }{|B_0|^2} ( 
\nabla \times b_0) \cdot a_0  - \frac{q_\perp^3}{3|B_0|^3}\,c_0 \cdot \nabla 
|B_0| + \frac{q_\perp^3}{2|B_0|^2}\, (\nabla \times a_0) \cdot b_0\,.
\end{aligned}
$$
This leads to
$$
 \delta \mu^{(2)} = \frac{q_\perp^2}{2|B_0|} \,\underbrace{\Big[ - \frac{1}{|B_0|} \Big( 
B_1 \cdot b_0 + 
\frac{1}{2}\,q_\para (\nabla \times b_0) \cdot b_0 \Big) + \frac{2q_\para}{|B_0|}\, 
\tn R \cdot b_0 \Big]}_{=: \sigma^*}\,,
 $$
where $\sigma^*$ is independent of $q_\perp$. Therefore, $\wh \mu = \mu(1+\eps\,\sigma^*)$, the map 
$\wh \tau:\wh \mu\mapsto q_\perp$ is given by $q_\perp = \sqrt{2\,\wh \mu
\,|B_0|/(1+\eps\,\sigma^*)}$ and the Hamiltonian correction reads 
$$
\delta H^{(2)} = -\wh \mu\,|B_0|\,\frac{\sigma^*}{1 + \eps\,\sigma^*}\,.
$$
The term $\cQ_2^*$ is given by
\be \label{Pgi2star}
 \begin{aligned}
 \cQ_2^* &= \frac{q_\perp}{2|B_0|} (\varrho_{2} \cdot \nabla B_0) 
\times a_0 
+ \frac{q_\perp}{2|B_0|} (a_0 \cdot \nabla 
B_0) \times \varrho_{2} + \frac{q_\perp^2}{2|B_0|^2} (a_0 \cdot \nabla B_1) \times a_0
 \\[2mm]
 &\quad - G^\para_{1}\, \frac{q_\perp}{|B_0|}\,a_0 \times (\nabla \times b_0) - G^\perp_{1}\, 
\frac{q_\perp}{|B_0|}\,a_0 \times 
(\nabla \times c_0) - G^\perp_1\,G^\theta_1\,a_0
 \\[3mm]
 &\quad - q_\para\,\varrho_{2} \times ( \nabla \times b_0) -  q_\perp\, \varrho_{2} \times ( \nabla 
\times c_0) 
 - q_\perp\,G^\theta_2\,a_0
 \\[2mm]
 &\quad - G^\theta_1\,\frac{q_\perp^2}{|B_0|}\,a_0 \cdot \nabla a_0 - 
\frac{q_\perp}{2}\,(G^\theta_1)^2\,c_0 + 
\frac{q_\perp^2}{|B_0|^2}\,G^\theta_1\, \nabla |B_0| - \frac{q_\perp^2}{6|B_0|^2}  a_0 
\times (a_0 \cdot \nabla)^2 B_0 
 \\[2mm]
 &\quad - \frac{q_\para}{2}\, \varrho_1 \times ( \nabla' 
\times (\varrho_1 \cdot \nabla b_0) )  - \frac{q_\perp}{2}\, \varrho_1 \times ( \nabla' 
\times (\varrho_1 \cdot \nabla c_0) ) 
 \\[2mm]
 &\quad + \frac{1}{2}\, \nabla B_0 \cdot ( \varrho_2 \times \varrho_1) - \frac{q_\perp}{|B_0|}\, 
\nabla |B_0| - q_\perp\, \nabla a_0 \cdot 
(\varrho_2 \times  b_0)
 \\[2mm]
 &\quad - \frac{q_\perp^3}{3|B_0|^3} (\nabla a_0 \times a_0) \cdot (a_0 \cdot \nabla B_0) 
-\frac{q_\perp^2}{2|B_0|^2} (\nabla a_0 \times a_0) \cdot B_1
 \\[2mm]
 &\quad -\frac{q_\para q_\perp^2}{2|B_0|^2} (\nabla a_0 \times a_0) \cdot (\nabla 
\times b_0) - \frac{q_\perp^3}{2|B_0|^2}  (\nabla a_0 \times a_0) \cdot (\nabla \times 
c_0) + \parfra{S_2^*}{r} \,,
\end{aligned}
\ee
where $\nabla'$ only acts on $\nabla b_0$ and $\nabla c_0$.
\end{lemma}

\section{Proofs} \label{sec:proofs}

\subsection{Proof of Proposition \ref{prop:1}} \label{proof:prop1}

The proof is split into three sections, with the following startegy in mind: first, for 
${(q_\tn{gy},\dot q_\tn{gy})}$ fixed, we consider the Lagrangian $L^\eps$ in \eqref{L:Y} as a 
function of $\eps$ 
and apply Taylor's theorem to write a series expansion in $\eps$ and estimate the remainder. In 
part two we compute the coefficients of this series expansion. This step involves a lot of 
algebra. Finally, we rewrite the series expansion of the Lagrangian so 
as to arrive at a gauge-invariant representation.

Let us introduce the following notation for the components of the GT \eqref{X(Y)}-\eqref{dotx(Y)},
\begin{align*}
 x_\eps(q_\tn{gy}) &:= \sum_{n=0}^{N+1} \eps^n\,\varrho_n(q_\tn{gy}) \,, \qquad\quad && \varrho_0 = 
r\,,
 \\
 v_{\para,\eps}(q_\tn{gy}) &:=  \sum_{n=0}^{N+1} \eps^n\,G^\para_n(q_\tn{gy})\,, \qquad\quad && 
G^\para_0 = 
q_\para\,,
 \\
 v_{\perp,\eps}(q_\tn{gy}) &:= \sum_{n=0}^{N+1} \eps^n\,G^\perp_n(q_\tn{gy})\,, \qquad\quad && 
G^\perp_0 = 
q_\perp\,,
 \\
 \theta_\eps(q_\tn{gy}) &:= \sum_{n=0}^{N+1} \eps^n\, G^\theta_n(q_\tn{gy}) \,, \qquad\quad && 
G^\theta_0 = 
\alpha\,,
 \\
 \dot x_\eps(q_\tn{gy},\dot q_\tn{gy}) &:= \sum_{n=0}^{N+1} \eps^n\,\dot \varrho_n(q_\tn{gy}, \dot 
 q_\tn{gy}) \,, 
\qquad\quad && \dot \varrho_0 = \dot r\,,
\end{align*}
where the time coordinate $t$ rests untransformed. For $(q_\tn{gy},\dot q_\tn{gy})$ fixed, we 
consider the Lagrangian \eqref{L:Y} as a function of 
$\eps$, split into three parts, $L^\eps = \vphi_0/\eps + \vphi_1 - \vphi_2\,\dot t$, with
\begin{subequations} \label{def:phi}
\begin{align}
 \vphi_0(\eps) &:= \dot x_\eps \cdot A_0(x_\eps)\,. \label{def:phi0}
 \\[3.5mm]
 \vphi_1(\eps) &:= v_{\para,\eps} \,\dot x_\eps \cdot b_0(x_\eps) + v_{\perp,\eps} \, \dot x_\eps 
\cdot c_0(x_\eps,\theta_\eps) + \dot x_\eps \cdot A_1(t,x_\eps) \,, \label{def:phi1}
 \\[1mm]
 \vphi_2(\eps) &:= \frac{v_{\para,\eps}^2}{2} + \frac{v_{\perp,\eps}^2}{2} + \phi (t, x_\eps ) \,. 
\label{def:phi2}
\end{align}
\end{subequations}
From assumption \ref{assumps} we have $A_0\in C^{N+3}(\Omega_x)$ and thus ${\vphi_0 \in 
C^{N+3}([0,\eps_\tn{max}])}$, since $\dot x_\eps$ is a polynomial in $\eps$. Also, 
$b_0,c_0,A_1,\phi \in 
C^{N+2}(\Omega_x)$ which implies $\vphi_1,\vphi_2 \in C^{N+2}([0,\eps_\tn{max}])$. Hence, we may 
apply Taylor's theorem and write
\be \label{Taylor}
\begin{aligned}
  \vphi_0(\eps) &= \sum_{j=0}^{N+1}\frac{\eps^j}{j!}\, \vphi_0^{(j)}(0) + O(\eps^{N+2})\,,
 \\[0mm]
 \vphi_1(\eps) &= \sum_{j=0}^{N}\frac{\eps^j}{j!}\, \vphi_1^{(j)}(0) + O(\eps^{N+1})\,,
 \\[0mm]
 \vphi_2(\eps) &= \sum_{j=0}^N\frac{\eps^j}{j!}\, \vphi_2^{(j)}(0) + O(\eps^{N+1})\,,
 \end{aligned}
\ee
where $\vphi^{(j)}$ denotes the $j$-th derivative of $\vphi$ with respect to $\eps$. The remainders 
are still $C^{1}$. This kind of regularity is necessary in the proof of Lemma 
\ref{lem:2} where, in order to apply the Gronwall's lemma 
\ref{Gronwall}, one needs the residual terms in the Euler-Lagrange equations to be continuous, 
which is guaranteed by the remainders being in $C^1$.

\subsubsection{Taylor coefficients}

Let us now compute 
the coefficients of the $\eps$-polynomials \eqref{Taylor}. For $j=0$ one has
\be \label{phi:lowest}
\begin{aligned}
  \vphi_0(0) &= A_0(r) \cdot \dot r \,,
  \\[3.5mm]
  \vphi_1(0) &= [ q_\para b_0(r) + q_\perp c_0(r) + A_1(t,r)] \cdot \dot r \,,
  \\[1mm]
  \vphi_2(0) &= \frac{q_\para^2}{2} + \frac{q_\perp^2}{2} + \phi(t,r) \,. 
\end{aligned}
\ee
To compute the derivatives of order $j$, we use the Leibniz rule,
 \be \label{Leibniz}
 (\vphi\,\chi)^{(j)} = \sum_{l=0}^{j} \binom{j}{l} \vphi^{(j-l)} \, \chi^{(l)}\,.
 \ee
For polynomials in $\eps$ we use the formula
 \be \label{polyder}
 \Big[ \sum_{n=0}^{N+1} \eps^n\,\varrho_n \Big]^{(j)} = \sum_{n=j}^{N+1} \frac{n!}{(n-j)!} 
\eps^{n-j}\,\varrho_n\,,
 \ee
 which leads to
 \be \label{polyder0}
  \Big[ \sum_{n=0}^{N+1} \eps^n\,\varrho_n \Big]^{(j)}(0) = j!\, \varrho_j\,.
 \ee
 We shall use the nable symbol to denote the gradient with respect to the position variable, 
$\nabla \equiv \pa/\pa x$. In order to write the Taylor expansion of a function $b_0(r + 
s(\eps,r))$ around $r$ we define 
the operator $\nabla'$, which acts only on the function $b_0$ and not on $s$. Hence,
 \begin{align*}
 b_0(r + s(\eps,r)) &= b_0(r) + \sum_i s_i\, \parfra{b_0(r)}{x_i} + \frac{1}{2} \sum_{i,j} 
s_i\,s_j\,\parfra{^2 b_0(r)}{x_i \pa x_j} + \frac{1}{6} \sum_{i,j,k} s_i\,s_j\,s_k 
\parfra{^3 b_0(r)}{x_i \pa x_j \pa x_k} + \ldots
 \\
 &= b_0(r) + s \cdot \nabla' b_0(r) + \frac{1}{2} (s \cdot \nabla')^2 b_0(r) + 
\frac{1}{6} (s \cdot \nabla')^3 b_0(r) + \ldots
 \end{align*}
 With the Leibniz rule \eqref{Leibniz}, for $j\geq1$ one computes
 \be \label{b0j:start}
 \begin{aligned}
 b_0^{(j)} (x_\eps) &= \Big( x_\eps^{(1)} \cdot \nabla' \,b_0 \Big)^{(j-1)} 
 \\[2mm]
 &= \sum_{k_1 = 0}^{j-1} 
\binom{j-1}{k_1} \Big( x_\eps^{(j-k_1)} \cdot \nabla' \Big)\, b_0^{(k_1)}
 \\[0mm]
 &= x_\eps^{(j)} \cdot \nabla'\, b_0 + \sum_{k_1 = 1}^{j-1} \binom{j-1}{k_1} \Big( x_\eps^{(j-k_1)} 
\cdot 
\nabla' \Big)\, b_0^{(k_1)}\,. 
\end{aligned}
\ee
Here and in the following, for sums we will use the convention 
\be \label{sumconv}
 \sum_{k=k_\tn{start}}^j \ldots = 0 \qquad \tn{if} \qquad j < k_\tn{start}\,.
\ee
Using the rule \eqref{b0j:start} two times yields
\begin{align*}
 b_0^{(j)} (x_\eps) &= x_\eps^{(j)} \cdot \nabla'\, b_0 + \sum_{k_1 = 1}^{j-1} \binom{j-1}{k_1} 
\Big( x_\eps^{(j-k_1)} 
\cdot 
\nabla' \Big)\, \Big( x_\eps^{(k_1)} \cdot \nabla' \Big)\, b_0
 \\
 &\quad + \sum_{k_1 = 2}^{j-1} \binom{j-1}{k_1} \Big( x_\eps^{(j-k_1)} 
\cdot \nabla' \Big)\, \sum_{k_2 =1}^{k_1-1} \binom{k_1-1}{k_2} \Big( x_\eps^{(k_1-k_2)} \cdot 
\nabla' \Big)\, b_0^{(k_2)}\,.
\end{align*}
We note that $x_\eps^{(k)}$ is a function of $\eps$ and $q_\tn{gy}$, and that $\nabla'$ 
only acts on $b_0$. Applying the rule \eqref{b0j:start} a third time leads to
\begin {align*}
 b_0^{(j)} (x_\eps) &= x_\eps^{(j)} \cdot \nabla'\, b_0 + \sum_{k_1 = 1}^{j-1} \binom{j-1}{k_1} 
\Big( x_\eps^{(j-k_1)} 
\cdot 
\nabla' \Big)\, \Big( x_\eps^{(k_1)} \cdot \nabla' \Big)\, b_0
 \\
 &\quad + \sum_{k_1 = 2}^{j-1} \binom{j-1}{k_1} \Big( x_\eps^{(j-k_1)} 
\cdot \nabla' \Big)\, \sum_{k_2 = 
1}^{k_1-1} \binom{k_1-1}{k_2} \Big( x_\eps^{(k_1-k_2)} \cdot \nabla' \Big)\, \Big( x_\eps^{(k_2)} 
\cdot \nabla' \Big)\, b_0
 \\[1mm]
 &\quad + \sum_{k_1 = 3}^{j-1} \binom{j-1}{k_1} \Big( x_\eps^{(j-k_1)} 
\cdot \nabla' \Big)\, \sum_{k_2 = 
2}^{k_1-1} \binom{k_1-1}{k_2} \Big( x_\eps^{(k_1-k_2)} \cdot \nabla' \Big)\, \, \sum_{k_3 = 
1}^{k_2-1} \binom{k_2-1}{k_3} \times
\\[1mm]
 & \qquad\qquad\qquad\qquad\qquad\qquad\qquad\qquad\qquad\qquad\qquad \times\Big( 
x_\eps^{(k_2-k_3)} \cdot \nabla' \Big)\, b_0^{(k_3)}\,.
\end{align*}
At each iteration, the starting index of sums in the last term gets raised by one due to 
the sum convention \eqref{sumconv}. One can thus apply the rule \eqref{b0j:start} $j-1$ times, 
until the last term becomes
\begin{align*}
 &\sum_{k_1 = j-1}^{j-1}  \, \sum_{k_2 = j-2}^{j-2} \ldots \, \sum_{k_{j-1} = 1}^{1} 
\binom{j-1}{k_1} \binom{k_1-1}{k_2} \ldots \binom{1}{k_{j-1}} \, \
 \\[2mm]
 & \qquad\quad \underbrace{\Big( x_\eps^{(j-k_1)} \cdot \nabla' \Big) \Big(x_\eps^{(k_1-k_2)} \cdot 
\nabla' 
\Big) \ldots \Big(x_\eps^{(k_{j-2}-k_{j-1})} \cdot \nabla' \Big)}_{j-1\:\mathrm{times}} \Big( 
x_\eps^{(k_{j-1})} \cdot 
\nabla' 
\Big)\, b_0 
 \\
 &= \Big( x_\eps^{(1)} \cdot \nabla' \Big)^j\, b_0 \,.
\end{align*}
Therefore, by applying the rule \eqref{b0j:start} recursively, for $j\geq1$ we can write the result 
in the compact form
\be \label{bj:der}
 b_0^{(j)} (x_\eps) = x_\eps^{(j)} \cdot \nabla b_0 + \cR_{j-1}^\eps \,,
\ee
with $\cR_{0}^\eps = 0$ and, for $j\geq 2$, with $\cR_{j-1}^\eps$ defined by
\be \label{bigTau}
\begin{aligned}
 \cR_{j-1}^\eps &:= \sum_{k_1 = 1}^{j-1} \binom{j-1}{k_1}  \Big( x_\eps^{(j-k_1)} \cdot 
\nabla' 
\Big) \Big( x_\eps^{(k_1)} \cdot \nabla' 
\Big) b_0
\\[2mm]
 &+ \sum_{k_1 = 2}^{j-1} \: \sum_{k_2 = 1}^{k_1-1}   \binom{j-1}{k_1} \binom{k_1-1}{k_2} 
  \Big( x_\eps^{(j-k_1)} \cdot \nabla' 
\Big) \Big( x_\eps^{(k_1-k_2)} \cdot \nabla' \Big) \Big( x_\eps^{(k_{2})} \cdot \nabla' 
\Big) b_0
\\[2mm]
 &+ \ldots
\\[2mm]
 &+ \sum_{k_1 = j-2\geq3}^{j-1} \: \sum_{k_2 = j-3}^{k_1-1} \ldots 
\sum_{k_{j-2} = 1 }^{k_{j-3}-1} \binom{j-1}{k_1} \binom{k_1-1}{k_2} \ldots 
\binom{k_{j-3}-1}{k_{j-2}}
\\[2mm]
&\qquad\quad  \Big( x_\eps^{(j-k_1)} \cdot \nabla' 
\Big) \Big( x_\eps^{(k_1-k_2)} \cdot \nabla' \Big) \ldots \Big( x_\eps^{(k_{j-3}-k_{j-2})} \cdot 
\nabla' 
\Big)\Big( x_\eps^{(k_{j-2})} \cdot \nabla' 
\Big) b_0
\\[2mm]
 &+ \Big( x_\eps^{(1)} \cdot \nabla' \Big)^j\,b_0 \,.
 \end{aligned}
 \ee
Since we need the derivatives \eqref{bj:der} evaluated at $\eps=0$, using \eqref{polyder0} 
leads to
 \be \label{b0j}
  b_0^{(j)} (x_\eps)\Big|_{\eps=0} = j! \Big[ \,\varrho_j \cdot \nabla 
b_0(r) + \cR_{j-1}(b_0) \Big]\,,\qquad j\geq 1\,,
 \ee
with $\cR_0 = 0$ and, for $j\geq 2$, with the term $\cR_{j-1}$ defined by
\begin{align}
 \cR_{j-1}(b_0) &:= \frac{1}{j!} \cT_{j-1}^0(b_0) = \sum_{k_1 = 1}^{j-1} \frac{(j-k_1)}{j}  \Big( 
\varrho_{j-k_1} \cdot \nabla' 
\Big) \Big( \varrho_{k_1} \cdot \nabla' 
\Big) b_0(r) \label{def:Rj-1}
\\[2mm]
 &+ \sum_{k_1 = 2}^{j-1} \: \sum_{k_2 = 1}^{k_1-1} \frac{(j-k_1)(k_1-k_2)}{jk_1} 
  \Big( \varrho_{j-k_1} \cdot \nabla' 
\Big) \Big( \varrho_{k_1-k_2} \cdot \nabla' \Big) \Big( \varrho_{k_2} \cdot \nabla' 
\Big) b_0(r) \nonumber
\\[2mm]
 &+ \ldots \nonumber
\\[2mm]
 &+ \sum_{k_1 = j-2\geq 3}^{j-1} \: \sum_{k_2 = j-3}^{k_1-1} \ldots 
\sum_{k_{j-2} = 1 }^{k_{j-3}-1} \frac{(j-k_1)(k_1-k_2) \ldots (k_{j-3}-k_{j-2})}{jk_1k_2 \ldots 
k_{j-3}} \nonumber
\\[2mm]
&\qquad\quad\qquad  \Big( \varrho_{j-k_1} \cdot \nabla' 
\Big) \Big( \varrho_{k_1-k_2} \cdot \nabla' \Big) \ldots \Big( \varrho_{k_{j-3}-k_{j-2}} \cdot 
\nabla' 
\Big)\Big( \varrho_{k_{j-2}} \cdot \nabla' 
\Big) b_0(r) \nonumber
\\[2mm]
& + \frac{1}{j!} \Big( \varrho_1 \cdot \nabla' \Big)^j\,b_0(r)\,. \nonumber
 \end{align} 
The computation of $c_0^{(j)}(x_\eps,\theta_\eps)$ is done in a similar way, i.e. we start as in 
\eqref{b0j:start},
\be \label{c0j:start}
 \begin{aligned}
 c_0^{(j)} (x_\eps,\theta_\eps) &= \Big( x_\eps^{(1)} \cdot \nabla' \,c_0 + \theta_\eps^{(1)} 
\parfra{}{\theta} c_0 \Big)^{(j-1)} 
 \\[2mm]
 &= \sum_{k_1 = 0}^{j-1} 
\binom{j-1}{k_1} \Big( x_\eps^{(j-k_1)} \cdot \nabla' + \theta_\eps^{(j-k_1)} \parfra{}{\theta} 
\Big)\, c_0^{(k_1)} 
 \\[0mm]
 &= x_\eps^{(j)} \cdot \nabla'\, c_0 + \theta_\eps^{(j)} \parfra{}{\theta} c_0 + \sum_{k_1 = 
1}^{j-1} \binom{j-1}{k_1} \Big( x_\eps^{(j-k_1)} \cdot \nabla' + \theta_\eps^{(j-k_1)} 
\parfra{}{\theta} \Big)\, c_0^{(k_1)}\,,
\end{aligned}
\ee
Applying this rule twice yields
 \begin{align*}
 c_0^{(j)} &= x_\eps^{(j)} \cdot \nabla'\, c_0 + \theta_\eps^{(j)} \parfra{}{\theta} c_0 + 
\sum_{k_1 = 1}^{j-1} 
\binom{j-1}{k_1} \Big( x_\eps^{(j-k_1)} \cdot \nabla' + \theta_\eps^{(j-k_1)} \parfra{}{\theta} 
\Big) 
 \\[0mm]
 & \qquad\qquad\qquad\qquad\qquad\qquad\qquad\qquad\quad \times \Big( x_\eps^{(k_1)} \cdot 
\nabla'\, c_0 + \theta_\eps^{(k_1)} \parfra{}{\theta} c_0 \Big)
 \\[1mm]
 & + \sum_{k_1 = 2}^{j-1} 
\binom{j-1}{k_1} \Big( x_\eps^{(j-k_1)} \cdot \nabla' + \theta_\eps^{(j-k_1)} \parfra{}{\theta} 
\Big) \sum_{k_2 = 1}^{k_1-1} 
\binom{k_1-1}{k_2} \Big( x_\eps^{(k_1-k_2)} \cdot \nabla' + \theta_\eps^{(k_1-k_2)} 
\parfra{}{\theta} 
\Big) c^{(k_2)}\,.
\end{align*}
We remind the reader again that $x_\eps^{(k)}$ as well as $\theta_\eps^{(k)}$ are functions of 
$\eps$ and $q_\tn{gy}$ and that $\nabla'$ and $\pa/\pa\theta$ 
only act on $c_0$. We can apply the rule \eqref{c0j:start} $j-1$ times, 
until the last term becomes
\begin{align*}
 &\sum_{k_1 = j-1}^{j-1}  \, \sum_{k_2 = j-2}^{j-2} \ldots \, \sum_{k_{j-1} = 1}^{1} 
\binom{j-1}{k_1} \binom{k_1-1}{k_2} \ldots \binom{1}{k_{j-1}} \, \
 \\[2mm]
 & \quad \times \Big( x_\eps^{(j-k_1)} \cdot \nabla' + \theta_\eps^{(j-k_1)} 
\parfra{}{\theta} 
\Big) \Big( x_\eps^{(k_1-k_2)} \cdot \nabla' + \theta_\eps^{(k_1-k_2)} 
\parfra{}{\theta} 
\Big) \ldots 
 \\[2mm]
 & \quad \times \Big(x_\eps^{(k_{j-2}-k_{j-1})} \cdot \nabla' + 
\theta_\eps^{(k_{j-2}-k_{j-1})} \parfra{}{\theta} \Big) \Big( x_\eps^{(k_{j-1})} \cdot 
\nabla' + \theta_\eps^{(k_{j-1})} \parfra{}{\theta} \Big)\, c_0 
 \\[2mm]
 &= \Big( x_\eps^{(1)} \cdot \nabla' + \theta_\eps^{(1)} \parfra{}{\theta} \Big)^j\, c_0 \,.
\end{align*}
Hence, with $\pa c_0/\pa\alpha = -a_0$, in analogy to \eqref{b0j} we arrive at
\be \label{c0j}
  c_0^{(j)} (x_\eps,\theta_\eps)\Big|_{\eps=0} = j! \Big[ \,\varrho_j \cdot \nabla 
c_0(r,\theta) - G^\theta_j\, a_0(r,\theta) + \cR^\alpha_{j-1}(c_0) \Big]\,,\qquad j\geq 1\,,
 \ee
with $\cR^\alpha_0 = 0$ and, for $j\geq 2$, with the term $\cR^\alpha_{j-1}$ defined by
\begin{align}
 \cR^\alpha_{j-1}(c_0) &:= \sum_{k_1 = 1}^{j-1} \frac{(j-k_1)}{j}  
\Big( \varrho_{j-k_1} \cdot \nabla' + G^\theta_{j-k_1} \parfra{}{\theta}
\Big) \Big( \varrho_{k_1} \cdot \nabla' + G^\theta_{k_1} \parfra{}{\theta}
\Big) c_0 \label{def:Rtheta}
\\[2mm]
 &+ \sum_{k_1 = 2}^{j-1} \: \sum_{k_2 = 1}^{k_1-1} \frac{(j-k_1)(k_1-k_2)}{jk_1} 
  \Big( \varrho_{j-k_1} \cdot \nabla' + G^\theta_{j-k_1} \parfra{}{\theta}
\Big) \Big( \varrho_{k_1-k_2} \cdot \nabla' + G^\theta_{k_1-k_2} \parfra{}{\theta} \Big) \nonumber
\\[2mm]
&\qquad\quad\qquad \times \Big( \varrho_{k_2} \cdot 
\nabla' + G^\theta_{k_2} \parfra{}{\theta} \Big) c_0 \nonumber
\\[2mm]
 &+ \ldots \nonumber
\\[2mm]
 &+ \sum_{k_1 = j-2\geq 3}^{j-1} \: \sum_{k_2 = j-3}^{k_1-1} \ldots 
\sum_{k_{j-2} = 1 }^{k_{j-3}-1} \frac{(j-k_1)(k_1-k_2) \ldots (k_{j-3}-k_{j-2})}{jk_1k_2 \ldots 
k_{j-3}} \nonumber
\\[2mm]
&\times \Big( \varrho_{j-k_1} \cdot \nabla' + G^\theta_{j-k_1} \parfra{}{\theta}
\Big) \Big( \varrho_{k_1-k_2} \cdot \nabla' + G^\theta_{k_1-k_2} 
\parfra{}{\theta}
\Big) \ldots \Big( \varrho_{k_{j-2}}  \cdot \nabla' + G^\theta_{k_{j-2}} \parfra{}{\theta}
\Big) c_0 \nonumber
\\[2mm]
& + \frac{1}{j!} \Big( \varrho_1 \cdot \nabla' + G^\theta_1 \parfra{}{\theta} \Big)^j\,c_0\,, 
\nonumber
 \end{align} 
where derivatives of $c_0$ are avaluated at $(r,\alpha)$. 

We have now all the material to compute 
the $j$-h derivative at $\eps=0$ 
of $\vphi_0$, $\vphi_1$ and $\vphi_1$, defined in 
\eqref{def:phi}, which are the coefficients of the Taylor 
expansions \eqref{Taylor}. Applying the Leibniz rule \eqref{Leibniz} twice, for the first term in 
\eqref{def:phi1} we obtain
\begin{align*}
 \Big[ v_{\para,\eps} \,\dot x_\eps \cdot b_0(x_\eps) \Big]^{(j)}(0) &= \bigg\{ 
\sum_{l=0}^j \binom{j}{l} \Big[ v_{\para,\eps} \,\dot x_\eps  \Big]^{(j-l)} \cdot 
b_0^{(l)}(x_\eps) \bigg\}_{\eps=0} 
 \\[1mm]
 & = \bigg\{ \sum_{l=0}^j 
\binom{j}{l} \Big[ \sum_{m=0}^{j-l} \binom{j-l}{m}\,v_{\para,\eps}^{(j-l-m)}\, \dot x_\eps^{(m)}  
\Big] \cdot b_0^{(l)}(x_\eps) \bigg\}_{\eps=0}\,. 
\end{align*}
Since $b_0^{(l)}(x_\eps)$ at $\eps=0$ is given by the formula \eqref{b0j} for $l\geq1$, we single 
out the summand with $l=0$, and insert \eqref{polyder0} to obtain, for $j\geq 1$,
\begin{align}
 &\Big[ v_{\para,\eps} \,\dot x_\eps \cdot b_0(x_\eps) \Big]^{(j)}(0) = \sum_{m=0}^{j} 
\frac{j!}{m!(j-m)!}\,(j-m)!\,G^\para_{j-m}\, m!\, \dot \varrho_m  \cdot 
b_0 \label{term1}
 \\
 &\:\: + \sum_{l=1}^j 
\frac{j!}{l!(j-l)!} \Big[ \sum_{m=0}^{j-l} \frac{(j-l)!}{m!(j-l-m)!} (j-l-m)!\, G^\para_{j-l-m}\, 
m!\, 
\dot \varrho_m  \Big] \cdot \,l! \Big[ \,\varrho_l \cdot \nabla b_0 + \cR_{l-1}(b_0) \Big]  
\nonumber
 \\
 & = j! \sum_{m=0}^{j} G^\para_{j-m}\, \dot \varrho_m  \cdot 
b_0 + j! \sum_{l=1}^j \Big[ \sum_{m=0}^{j-l} G^\para_{j-l-m}\, 
\dot \varrho_m  \Big] \cdot \, \Big[ \,\varrho_l \cdot \nabla b_0 + \cR_{l-1}(b_0) \Big]\,,  
\nonumber
\end{align}
With the same reasoning, using the result \eqref{c0j}, for the 
second term in \eqref{def:phi1} one obtains 
\be \label{term2}
\begin{aligned}
 &\Big[ v_{\perp,\eps} \,\dot x_\eps \cdot c_0(x_\eps,\theta_\eps) \Big]^{(j)}(0)
 \\
 & \qquad = j! \sum_{m=0}^{j} G^\perp_{j-m}\, \dot \varrho_m  \cdot 
c_0 + j! \sum_{l=1}^j \Big[ \sum_{m=0}^{j-l} G^\perp_{j-l-m}\, 
\dot \varrho_m  \Big] \cdot \, \Big[ \,\varrho_l \cdot \nabla c_0 - G^\theta_l\,a_0 + 
\cR^\alpha_{l-1}(c_0) 
\Big]\,, 
\end{aligned}
\ee
For the third term in \eqref{def:phi1} we apply the Leibniz rule \eqref{Leibniz} once to compute
\be \label{term3}
\begin{aligned}
 \Big[ \dot x_\eps \cdot A_1(t,x_\eps) \Big]^{(j)}(0) &= \bigg\{ \sum_{l=0}^j 
\binom{j}{l} \dot x_\eps^{(j-l)} \cdot A_1^{(l)}(t,x_\eps) \bigg\}_{\eps=0}
 \\
 & = j!\, \dot \varrho_j \cdot A_1 + j!\sum_{l=1}^j \dot \varrho_{j-l} \cdot \Big[ 
\,\varrho_l \cdot 
\nabla A_1 + \cR_{l-1}(A_1) \Big]\,.
\end{aligned}
\ee
From this result we can also compute the derivatives of $\vphi_0$ given in \eqref{def:phi0},
\be \label{phi0j}
\begin{aligned}
 \frac{\vphi_0^{(j)}(0)}{j!} &=  \dot \varrho_j \cdot A_0 + 
\sum_{l=1}^j \dot \varrho_{j-l} \cdot \Big[ 
\,\varrho_l \cdot 
\nabla A_0 + \cR_{l-1}(A_0) \Big]\,.
\end{aligned}
\ee
For $\vphi_1$ given in \eqref{def:phi1}, using also the previous result \eqref{term1}, we finally 
obtain
\be \label{phi1j}
\begin{aligned}
 \frac{\vphi_1^{(j)}(0)}{j!} &=  \sum_{m=0}^{j} G^\para_{j-m}\, \dot \varrho_m  \cdot 
b_0 +  \sum_{l=1}^j \Big[ \sum_{m=0}^{j-l} G^\para_{j-l-m}\, 
\dot \varrho_m  \Big] \cdot  \Big[ \varrho_l \cdot \nabla b_0 + \cR_{l-1}(b_0) \Big]
 \\[1mm]
  & +   \sum_{m=0}^{j} G^\perp_{j-m}\, \dot \varrho_m  \cdot 
c_0 +  \sum_{l=1}^j \Big[ \sum_{m=0}^{j-l} G^\perp_{j-l-m}\, 
\dot \varrho_m  \Big] \cdot  \Big[ \varrho_l \cdot \nabla c_0 - G^\theta_l\,a_0 + 
\cR_{l-1}^\alpha(c_0) 
\Big]
 \\[1mm]
  & +  \dot \varrho_j \cdot A_1 + \sum_{l=1}^j \dot \varrho_{j-l} \cdot \Big[ 
\,\varrho_l \cdot 
\nabla A_1 + \cR_{l-1}(A_1) \Big] \,.
\end{aligned}
\ee
For $\vphi_2$ given in \eqref{def:phi2}, using \eqref{polyder0} and \eqref{b0j}, one obtains
\be \label{phi2j}
 \frac{\vphi_2^{(j)}(0)}{j!} = q_\para G^\para_j + q_\perp G^\perp_j + \varrho_j \cdot \nabla 
\phi(r) + 
\cR_{j-1}(\phi)\,.
\ee

We shall transform the expressions \eqref{phi0j}-\eqref{phi1j} a bit further. 
In particular, in the sum over $l$ in \eqref{phi0j} and in the last line of \eqref{phi1j} we 
single out the term with $l=j$. Additionally, in the sums over $m$ in \eqref{phi1j} we single out 
the terms with $m=0$. For $j\geq 1$, this leads to
\be \label{phi0jB}
\begin{aligned}
 \frac{\vphi_0^{(j)}(0)}{j!} &=  \dot \varrho_j \cdot A_0 + \dot r \cdot \Big[ 
\,\varrho_j \cdot \nabla A_0 + \cR_{j-1}(A_0) \Big] + 
\sum_{l=1}^{j-1} \dot \varrho_{j-l} \cdot \Big[ 
\,\varrho_l \cdot \nabla A_0 + \cR_{l-1}(A_0) \Big] 
\end{aligned}
\ee
and to
\be \label{phi1jB}
\begin{aligned}
 & \frac{\vphi_1^{(j)}(0)}{j!} =  G^\para_j\,\dot r \cdot b_0 + \sum_{m=1}^{j} G^\para_{j-m}\, \dot 
\varrho_m  \cdot 
b_0 +  \sum_{l=1}^j  G^\para_{j-l}\,\Big[ \varrho_l \cdot \nabla b_0 + \cR_{l-1}(b_0) \Big]\cdot 
\dot r
 \\
 & +  \sum_{l=1}^{j-1} \Big[ \sum_{m=1}^{j-l} G^\para_{j-l-m}\, 
\dot \varrho_m  \Big] \cdot  \Big[ \varrho_l \cdot \nabla b_0 + \cR_{l-1}(b_0) \Big]
 \\
  & + G^\perp_j\,\dot r \cdot c_0 + \sum_{m=1}^{j} G^\perp_{j-m}\, \dot \varrho_m  \cdot 
c_0 +  \sum_{l=1}^j G^\perp_{j-l}\,  \Big[ \varrho_l \cdot \nabla c_0 - G^\theta_l\,a_0 + 
\cR_{l-1}^\alpha(c_0) \Big] \cdot \dot r
 \\
 & +  \sum_{l=1}^{j-1} \Big[ \sum_{m=1}^{j-l} G^\perp_{j-l-m}\, 
\dot \varrho_m  \Big] \cdot  \Big[ \varrho_l \cdot \nabla c_0 - G^\theta_l\,a_0 + 
\cR_{l-1}^\alpha(c_0) \Big]
 \\
  & +  \dot \varrho_j \cdot A_1 + \dot r \cdot \Big[ 
\,\varrho_j \cdot \nabla A_1 + \cR_{j-1}(A_1) \Big] + \sum_{l=1}^{j-1} \dot \varrho_{j-l} \cdot 
\Big[ 
\,\varrho_l \cdot \nabla A_1 + \cR_{l-1}(A_1) \Big] \,.
\end{aligned}
\ee
It will be convenient to eliminate the terms $\dot \varrho_m$ from the first and the third line of 
\eqref{phi1jB}, using the equivalence of Lagrangians from definition \eqref{def:equiv}:
\be \label{bitfurther2}
\begin{aligned}
 &\sum_{m=1}^{j} G^\para_{j-m}\, \dot \varrho_m  \cdot b_0 + \sum_{l=1}^{j} G^\para_{j-l} 
\,\varrho_l \cdot 
\nabla 
b_0 \cdot \dot r
 \\
 &= \sum_{m=1}^{j} \Big[ \ds{} (G^\para_{j-m} \,\varrho_m  \cdot b_0) - \dot G^\para_{j-m} 
(\varrho_m \cdot 
b_0) 
- 
G^\para_{j-m}\,\dot r \cdot (\varrho_m \times (\nabla \times b_0)) \Big]\,,
\end{aligned}
\ee
and, respectively,
\be \label{bitfurther1}
\begin{aligned}
 &\sum_{m=1}^{j} G^\perp_{j-m}\, \dot \varrho_m  \cdot c_0 + \sum_{l=1}^{j} G^\perp_{j-l} 
\,(\varrho_l \cdot 
\nabla 
c_0 - G^\theta_l\,a_0 ) \cdot \dot r
 \\
 &= \sum_{m=1}^{j} G^\perp_{j-m} [ \dot \varrho_m  \cdot c_0 + ( \varrho_m \cdot 
\nabla c_0 - G^\theta_m\,a_0 ) \cdot \dot r ]
 \\
 &= \sum_{m=1}^{j} G^\perp_{j-m} \Big[ \ds{} (\varrho_m  \cdot c_0) - (\dot r \cdot \nabla c_0 - 
\dot 
\alpha\,a_0) \cdot \varrho_m +  (\varrho_m \cdot \nabla c_0 - G^\theta_m\,a_0) \cdot \dot r \Big]
 \\
 &= \sum_{m=1}^{j} \Big[ \ds{} (G^\perp_{j-m}\, \varrho_m  \cdot c_0) - \dot 
G^\perp_{j-m}(\varrho_m \cdot c_0) 
- 
G^\perp_{j-m}\,\dot r \cdot (\varrho_m \times (\nabla \times c_0)) \Big.
 \\
 &\Big. \qquad\quad + G^\perp_{j-m}\,\dot\alpha\, (\varrho_m \cdot a_0)  - 
G^\perp_{j-m}\,G^\theta_m\,a_0 \cdot \dot r 
\Big]\,,
\end{aligned}
\ee
which leads to
\begin{align}
 & \frac{\vphi_1^{(j)}(0)}{j!} \sim  G^\para_j\,\dot r \cdot b_0 + G^\perp_j\,\dot r 
\cdot c_0 - \sum_{m=1}^{j} \Big[ \dot G^\para_{j-m} 
(\varrho_m \cdot b_0) + G^\para_{j-m}\,\dot r \cdot (\varrho_m \times (\nabla \times b_0)) \Big]  
\label{phi1jC}
 \\
 & +  \sum_{l=1}^{j-1} \Big[\sum_{m=1}^{j-l} G^\para_{j-l-m}\, \dot \varrho_m  \Big] \cdot  \Big[ 
\varrho_l 
\cdot \nabla 
b_0 + \cR_{l-1}(b_0) \Big] +  \sum_{l=2}^j  G^\para_{j-l}\, \cR_{l-1}(b_0) \cdot \dot r    \nonumber
 \\
  &  - \sum_{m=1}^{j} \Big[ \dot G^\perp_{j-m}(\varrho_m \cdot c_0)  + 
G^\perp_{j-m}\,\dot r \cdot (\varrho_m \times (\nabla \times c_0)) - G^\perp_{j-m}\,\dot\alpha\, 
(\varrho_m 
\cdot a_0)  + 
G^\perp_{j-m}\,G^\theta_m\,a_0 \cdot \dot r  \Big]  \nonumber
  \\
 & + \sum_{l=1}^{j-1} \Big[ \sum_{m=1}^{j-l} G^\perp_{j-l-m}\, 
 \dot \varrho_m  \Big] \cdot  \Big[ \varrho_l \cdot \nabla c_0 - G^\theta_l\,a_0 + 
\cR_{l-1}^\alpha(c_0) 
\Big]  + 
\sum_{l=2}^j  G^\perp_{j-l}\,\cR_{l-1}^\alpha(c_0) \cdot \dot r  
\nonumber
 \\[0mm]
  & +  \dot \varrho_j \cdot A_1 + \dot r \cdot \Big[ 
\,\varrho_j \cdot \nabla A_1 + \cR_{j-1}(A_1) \Big] + \sum_{l=1}^{j-1} \dot \varrho_{j-l} \cdot 
\Big[ 
\,\varrho_l \cdot \nabla A_1 + \cR_{l-1}(A_1) \Big] \,.  \nonumber
\end{align}
The Lagrangian $L^\eps$ now reads
\begin{align}
 L^\eps &= \frac{\vphi_0(\eps)}{\eps} + \vphi_1(\eps) - \vphi_2(\eps) \,\dot t 
\label{Leps}
 \\[1mm]
 &= \sum_{j=0}^{N+1} \frac{\eps^{j-1}}{j!}\, \vphi_0^{(j)}(0) + \sum_{j=0}^{N} 
\frac{\eps^{j}}{j!}\, \vphi_1^{(j)}(0) - \sum_{j=0}^{N} 
\frac{\eps^{j}}{j!}\, \vphi_2^{(j)}(0)\,\dot t + O(\eps^{N+1})   \nonumber
 \\
 &= \frac{\vphi_0(0)}{\eps} + \vphi_0^{(1)}(0) + \vphi_1(0) - \vphi_2(0) \,\dot t + 
\sum_{j=1}^{N} 
\eps^j \Big[ \frac{\vphi_0^{(j+1)}(0)}{(j+1)!} +
\frac{\vphi_1^{(j)}(0)}{(j)!} - \frac{\vphi_2^{(j)}(0)}{(j)!}\,\dot t \Big] + O(\eps^{N+1}) \,,  
\nonumber
\end{align}
where we used \eqref{Taylor} to estimate the remainder.
The terms $\vphi_0(0)$, $\vphi_1(0)$ and $\vphi_2(0)$ are given in \eqref{phi:lowest}. From 
\eqref{phi0jB} one computes
\be \label{sumterm0}
 \vphi_0^{(1)}(0) = \dot \varrho_1 \cdot A_0 + \varrho_1 \cdot \nabla A_0 \cdot \dot r \,.
\ee
For $j \geq 1$, from \eqref{phi2j},\eqref{phi0jB} and \eqref{phi1jC} one computes 
\be \label{sumterms}
\begin{aligned}
 &\frac{\vphi_0^{(j+1)}(0)}{(j+1)!} + \frac{\vphi_1^{(j)}(0)}{(j)!} - 
\frac{\vphi_2^{(j)}(0)}{(j)!}\,\dot t 
 = \dot \varrho_{j+1} \cdot A_0 + \varrho_{j+1} \cdot 
\nabla A_0 \cdot \dot r + \dot \varrho_{j} \cdot A_1  + \varrho_{j} \cdot 
\nabla A_1 \cdot \dot r 
 \\[2mm]
 &\qquad\quad +  \Big[ G^\para_{j} b_0 + G^\perp_{j} c_0 + \cQ_{j}^*(q_\tn{gy}) \Big] 
\cdot \dot r  - \Big[ q_\para G^\para_{j} + q_\perp G^\perp_{j} + \varrho_{j} \cdot \nabla \phi 
\Big]\,\dot t  + \cL_{j}^*(q_\tn{gy}, \dot q_\tn{gy}) \,,
\end{aligned}
\ee
with
\be \label{def:Pj-1}
\begin{aligned}
 \cQ_{j}^*(q_\tn{gy}) &:= \cR_{j}(A_0) - \sum_{m=1}^{j} \Big[ G^\para_{j-m}\, (\varrho_m \times 
(\nabla \times 
b_0)) + G^\perp_{j-m}\, (\varrho_m \times (\nabla \times 
c_0))\Big]
 \\
 &\: + \cR_{j-1}(A_1) +  \sum_{l=2}^j \Big[ G^\para_{j-l}\, \cR_{l-1}(b_0) +  
G^\perp_{j-l}\,\cR_{l-1}^\alpha(c_0) \Big] - \sum_{m=1}^{j}  G^\perp_{j-m}\,G^\theta_m\,a_0  \,,
\end{aligned}
\ee
and
\be \label{def:Qj-1}
\begin{aligned}
 \cL_{j}^*(q_\tn{gy}, \dot q_\tn{gy}) &:= \sum_{l=1}^{j} \dot \varrho_{j+1-l} \cdot \Big[ 
\,\varrho_l \cdot \nabla A_0 + \cR_{l-1}(A_0) \Big] 
 \\
 &\:\, - \sum_{m=1}^{j} \Big[ \dot G^\para_{j-m} 
(\varrho_m \cdot b_0) + \dot G^\perp_{j-m} (\varrho_m \cdot c_0) - G^\perp_{j-m}\,\dot\alpha\, 
(\varrho_m \cdot 
a_0)  \Big]  
 \\
 &\:\, +  \sum_{l=1}^{j-1} \Big[\sum_{m=1}^{j-l} G^\para_{j-l-m}\, \dot \varrho_m  \Big] \cdot  
\Big[ 
\varrho_l \cdot 
\nabla b_0 + \cR_{l-1}(b_0) \Big] 
 \\
 &\:\, + \sum_{l=1}^{j-1} \Big[ \sum_{m=1}^{j-l} G^\perp_{j-l-m}\, 
 \dot \varrho_m  \Big] \cdot  \Big[ \varrho_l \cdot \nabla c_0 - G^\theta_l\,a_0 + 
\cR_{l-1}^\alpha(c_0) 
\Big]
 \\
 &\:\, + \sum_{l=1}^{j-1} \dot \varrho_{j-l} \cdot \Big[ 
\,\varrho_l \cdot \nabla A_1 + \cR_{l-1}(A_1) \Big] - \cR_{j-1}(\phi)\,\dot t\,.
\end{aligned}
\ee
Here, $\cR$ and $\cR^\alpha$ have been defined in \eqref{def:Rj-1} and \eqref{def:Rtheta}, 
respectively, and we remind the reader of the sum convention \eqref{sumconv}.

\subsubsection{Gauge-invariant formulation}

It will be convenient to write \eqref{sumterm0} and \eqref{sumterms} in terms of the 
electromagnetic fields $E,B$ rather than the potentials $\phi,A$. For this, consider 
the product rule on the tangent space,
$$
 \ds{}(\varrho_j \cdot A) = \dot \varrho_j \cdot A + \varrho_j \cdot  \dot A = \dot \varrho_j \cdot 
A +   
 (\dot r \cdot \nabla A + \dot t\,\pa_t A) \cdot \varrho_j \,.
$$
where we used the tangent map to express $\dot A$. One can thus use the identity
\be \label{usealot}
{\nabla A \cdot \varrho_j - \varrho_j \cdot \nabla A = \varrho_j 
\times (\nabla \times 
A)} = \varrho_j \times B
\ee
to obtain
$$
 \dot \varrho_j \cdot A + \varrho_j \cdot \nabla A \cdot \dot r = \ds{} (\varrho_j \cdot A) - \dot 
r \cdot ( \varrho_j \times B) - \dot t\,\pa_t A \cdot \varrho_j \,.
$$
Therefore, by definition \eqref{def:equiv}, one has the equivalence
\be \label{fields1}
\begin{aligned}
 \dot \varrho_j \cdot A_0 + \varrho_j \cdot \nabla A_0 \cdot \dot r &\:\: \sim \: - \dot r \cdot ( 
\varrho_j \times 
B_0)
 \\[3mm]
 \dot \varrho_j \cdot A_1 + \varrho_j \cdot \nabla A_1 \cdot \dot r - \varrho_j \cdot \nabla \phi 
\,\dot t
&\:\: \sim \: - 
\dot r \cdot ( \varrho_j \times 
B_1) + \varrho_j \cdot E\,\dot t\,.
\end{aligned}
\ee
The only remaining terms featuring the 
electromagnetic potentials are the first terms in each line of \eqref{def:Pj-1}, as well as the 
first and the last line of \eqref{def:Qj-1}. Since $\cR$ is linear, these terms are of the generic 
form
\be \label{Qstart}
 \cR_{j-1}(A \cdot \dot r - \phi\,\dot t) + \sum_{l=1}^{j-1} \dot \varrho_{j-l} \cdot \Big[ 
\,\varrho_l \cdot \nabla A + \cR_{l-1}(A) \Big] \,.
\ee
From \eqref{def:Rj-1} we write
\be \label{newR}
\begin{aligned}
 \cR_{j-1}(A \cdot \dot r - \phi\,\dot t) &= \sum_{k_1=1}^{j-1} \frac{j-k_1}{j} (\varrho_{j-k_1} 
\cdot 
\nabla')(\varrho_{k_1} 
\cdot \nabla') (A \cdot \dot r - \phi\,\dot t) + \cR_{j-1}^\tn{I}(A \cdot \dot r - \phi\,\dot t)\,,
\end{aligned}
\ee
where for $j\geq 3$ we defined 
\begin{align}
 \cR_{j-1}^\tn{I} &(A \cdot \dot r - \phi\,\dot t) := \sum_{k_1 = 2}^{j-1} \: \sum_{k_2 = 
1}^{k_1-1} 
\frac{(j-k_1)(k_1-k_2)}{jk_1} 
   \nonumber
\\[2mm]
 & \qquad\qquad\qquad \times \Big( \varrho_{j-k_1} \cdot \nabla' 
\Big) \Big( \varrho_{k_1-k_2} \cdot \nabla' \Big) \Big( \varrho_{k_2} \cdot \nabla' 
\Big) (A \cdot \dot r - \phi\,\dot t) + \ldots \label{def:RjI}
\\[2mm]
 &+ \sum_{k_1 = j-2\geq 3}^{j-1} \: \sum_{k_2 = j-3}^{k_1-1} \ldots 
\sum_{k_{j-2} = 1 }^{k_{j-3}-1} \frac{(j-k_1)(k_1-k_2) \ldots 
(k_{j-3}-k_{j-3})}{jk_1k_2 \ldots 
k_{j-3}} \nonumber
\\[2mm]
&\qquad\quad\qquad  \Big( \varrho_{j-k_1} \cdot \nabla' 
\Big) \Big( \varrho_{k_1-k_2} \cdot \nabla' \Big) \ldots \Big( \varrho_{k_{j-3}-k_{j-2}} \cdot 
\nabla' 
\Big)\Big( \varrho_{k_{j-2}} \cdot \nabla' 
\Big) (A \cdot \dot r - \phi\,\dot t) \nonumber
\\[2mm]
& + \frac{1}{j!} \Big( \varrho_1 \cdot \nabla' \Big)^{j}\,(A \cdot \dot r - \phi\,\dot t)\,. 
\nonumber
 \end{align} 
The first term in the sum of \eqref{Qstart} can be written as
\begin{align}
 &\sum_{l=1}^{j-1} \varrho_l \cdot \nabla A \cdot \dot \varrho_{j-l} = \frac{1}{2} \sum_{l=1}^{j} ( 
\varrho_l \cdot 
\nabla A \cdot \dot \varrho_{j-l} + \varrho_{j-l} \cdot 
\nabla A \cdot \dot \varrho_{l} )   \label{HO:1}
 \\[0mm]
 &\qquad\quad= \frac{1}{2} \sum_{l=1}^{j-1} ( \varrho_l \cdot 
\nabla A \cdot \dot \varrho_{j-l} - \dot \varrho_{j-l} \cdot 
\nabla A \cdot \varrho_{l} )  \nonumber
 \\[0mm]
 &\qquad\quad\quad + \frac{1}{2} \sum_{l=1}^{j-1} \Big[  \ds{} (\varrho_{j-l} \cdot 
\nabla A \cdot  \varrho_{l} )  - \varrho_{j-l} \cdot
\nabla (\dot r \cdot  \nabla A + \dot t\,\pa_t A) \cdot \varrho_{l} \Big]   \nonumber
 \\[0mm]
 &\qquad\quad = \frac{1}{2} \sum_{l=1}^{j-1} \Big[ - \dot \varrho_{j-l} \cdot (\varrho_l \times 
B_0) 
+  \ds{} 
(\varrho_{j-l} \cdot 
\nabla A \cdot  \varrho_{l} ) - \varrho_{j-l} \cdot
\nabla (\dot r \cdot  \nabla A + \dot t\,\pa_t A) \cdot \varrho_{l} \Big]\,.   \nonumber
\end{align}
Moreover, in \eqref{newR},
$$
\begin{aligned}
 &\sum_{l=1}^{j-1} \frac{j-l}{j} (\varrho_{j-l} \cdot \nabla')(\varrho_l \cdot \nabla') (A \cdot 
\dot r - \phi\,\dot t)
 \\
 = \frac{1}{2} &\sum_{l=1}^{j-1} \Big[ \frac{j-l}{j} (\varrho_{j-l} 
\cdot \nabla')(\varrho_{l} \cdot \nabla') +  \frac{l}{j} (\varrho_{l} \cdot \nabla')(\varrho_{j-l} 
\cdot 
\nabla') \Big] (A \cdot \dot r - \phi\,\dot t)
 \\
 = \frac{1}{2} &\sum_{l=1}^{j-1} (\varrho_{j-l} 
\cdot \nabla')(\varrho_{l} \cdot \nabla') (A \cdot \dot r - \phi\,\dot t)  \,.
\end{aligned}
$$
Combining this result with the last term in the last line of \eqref{HO:1} yields
$$
\begin{aligned}
 \frac{1}{2} &\sum_{l=1}^{j-1} \Big[ (\varrho_{j-l} \cdot \nabla')(\varrho_{l} \cdot \nabla') (A 
\cdot \dot r - \phi\,\dot t) - 
 \varrho_{j-l} \cdot \nabla' (\dot r \cdot  \nabla A + \dot t\,\pa_t A) \cdot \varrho_{l}  \Big] 
 \\
 = \frac{1}{2} &\sum_{l=1}^{j-1} \Big[ (\varrho_{j-l} \cdot \nabla') (\varrho_{l} \cdot \nabla A   
- 
 \nabla A \cdot \varrho_{l}) \cdot \dot r  \Big] - \frac{1}{2} \sum_{l=1}^{j-1} \varrho_{j-l} \cdot 
\nabla'( \nabla\phi\,\dot t + \dot t\,\pa_t A) \cdot \varrho_{l}
 \\
 = \frac{1}{2} &\sum_{l=1}^{j-1} \Big[  (\varrho_{j-l} \cdot \nabla B_0) \times \varrho_{l}  \Big] 
\cdot \dot 
r + \frac{1}{2} \sum_{l=1}^{j-1} \varrho_{j-l} \cdot \nabla E \cdot \varrho_{l}\,\dot t \,.
\end{aligned}
$$
Hence we arrived at
\begin{align}
 &\cR_{j-1}(A \cdot \dot r - \phi\,\dot t)  + \sum_{l=1}^{j-1} \dot \varrho_{j-l} \cdot \Big[ 
\,\varrho_l \cdot \nabla A + \cR_{l-1}(A) \Big] \label{almostthere}
 \\[0mm]
 & \: \sim\: \frac{1}{2} \sum_{l=1}^{j-1} \Big[  
(\varrho_{j-l} \cdot \nabla B) \times \varrho_{l}  \Big] \cdot \dot r + \frac{1}{2} 
\sum_{l=1}^{j-1} 
\varrho_{j-l} \cdot \nabla E \cdot \varrho_{l}\,\dot t - \frac{1}{2} \sum_{l=1}^{j-1} 
\dot \varrho_{j-l} \cdot (\varrho_l \times B)  \nonumber
 \\[0mm]
 &\: \quad + \cR_{j-1}^\tn{I}(A \cdot \dot r - \phi\,\dot t) + \sum_{l=1}^{j-1} \dot 
\varrho_{j-l} \cdot \cR_{l-1}(A) \,.  \nonumber
\end{align}
The terms in the last line still contain the electromagentic potentials instead of the fields. We 
were not able to prove that a field representation exists at all orders. However, we can easily 
prove it for $j=3$:
\begin{align}
  &\cR_2^\tn{I}(A \cdot \dot r - \phi\,\dot t) + 
\dot \varrho_{1} \cdot \cR_{1}(A)   \label{HO:id1}
 \\[2mm]
 &\quad = \frac{1}{3!}(\varrho_1 \cdot \nabla')^3 (A \cdot \dot r - \phi\,\dot t) + 
\frac{1}{2!}(\varrho_1 \cdot \nabla')^2 A \cdot \dot \varrho_1   \nonumber
 \\[2mm]
 &\quad = \frac{1}{6}(\varrho_1 \cdot \nabla')^2 ( \varrho_1 \cdot \nabla (A \cdot \dot r - 
\phi\,\dot t) + 
A \cdot 
\dot 
\varrho_1) + \frac{1}{3}(\varrho_1 \cdot \nabla')^2 A \cdot \dot \varrho_1   \nonumber
 \\[2mm]
 &\quad = \frac{1}{6}(\varrho_1 \cdot \nabla')^2 \Big[ \varrho_1 \cdot \nabla (A \cdot \dot r - 
\phi\,\dot t) + \ds{} 
(A \cdot \varrho_1) - (\dot r \cdot \nabla A + \dot t\,\pa_t A) \cdot \varrho_1 \Big] + 
\frac{1}{3}(\varrho_1 
\cdot \nabla')^2 A \cdot \dot \varrho_1  \nonumber
 \\[2mm]
 &\quad = - \frac{1}{6}\, \dot r \cdot [ \varrho_1 \times (\varrho_1 \cdot \nabla')^2 B ] + 
\frac{1}{6} (\varrho_1 
\cdot \nabla') (\varrho_1 \cdot \nabla E \cdot \varrho_1)\,\dot t + 
\frac{1}{6} \ds{} [(\varrho_1 \cdot \nabla')^2 A \cdot \varrho_1]  \nonumber
 \\[2mm]
 &\quad \qquad - \frac{1}{3} (\varrho_1 \cdot \nabla') (\dot \varrho_1 \cdot \nabla A \cdot 
\varrho_1)
 + \frac{1}{3}(\varrho_1 \cdot \nabla')^2 A \cdot \dot \varrho_1   \nonumber
 \\[2mm]
 &\quad \sim - \frac{1}{6} \dot r \cdot [ \varrho_1 \times (\varrho_1 \cdot \nabla')^2 B ] + 
\frac{1}{6} (\varrho_1 
\cdot \nabla') (\varrho_1 \cdot \nabla E \cdot \varrho_1)\,\dot t - 
\frac{1}{3}\, \dot \varrho_1 \cdot [ \varrho_1 \times (\varrho_1 \cdot \nabla B) ]\,.  \nonumber
\end{align}
We conjecture that such field representations can be derived at every order and leave the proof 
for later. 

In summary, the above algebra leads to the following representation of \eqref{sumterm0} 
and \eqref{sumterms}, respectively,
$$
 \vphi_0^{(1)}(0) \sim - \dot r \cdot ( \varrho_1 \times B_0) \,,
$$
and, for $n \geq 1$ (switching the index $j$ to $n$),
$$
\begin{aligned}
 &\frac{\vphi_0^{(n+1)}(0)}{(n+1)!} + \frac{\vphi_1^{(n)}(0)}{(n)!} - 
\frac{\vphi_2^{(n)}(0)}{(n)!}\,\dot t 
 \sim  \Big[ G^\para_{n} b_0 + G^\perp_{n} c_0 - \varrho_{n+1} \times B_0 - \varrho_n \times B_1 + 
\cQ_{n} (q_\tn{gy}) \Big] \cdot \dot r
 \\[2mm]
 &\qquad\qquad\quad - (\varrho_n \cdot b_0)\, \dot q_\para - 
G_{n-1}^\theta\,q_\perp\,(\dot \varrho_1 \cdot a_0) - \Big[ q_\para G^\para_{n} + q_\perp 
G^\perp_{n} - \varrho_{n} \cdot E 
\Big]\,\dot t + 
\cL_{n}(q_\tn{gy}, \dot q_\tn{gy}) \,,
\end{aligned}
$$
with
\be \label{def:cQgy}
\begin{aligned}
 \cQ_{n}(q_\tn{gy}) &:= \frac{1}{2} \sum_{l=1}^{n} 
(\varrho_{n+1-l} \cdot \nabla B_0) \times \varrho_{l} + \frac{1}{2} \sum_{l=1}^{n-1}  
(\varrho_{n-l} \cdot \nabla B_1) \times \varrho_{l}
 \\
 &\: - \sum_{m=1}^{n} \Big[ G^\para_{n-m}\, (\varrho_m \times (\nabla \times 
b_0)) + G^\perp_{n-m}\, (\varrho_m \times (\nabla \times 
c_0))\Big]
 \\
 &\: +  \sum_{l=2}^n \Big[ G^\para_{n-l}\, \cR_{l-1}(b_0) +  
G^\perp_{n-l}\,\cR_{l-1}^\alpha(c_0) \Big] - \sum_{m=1}^{n}  G^\perp_{n-m}\,G^\theta_m\,a_0  \,,
\end{aligned}
\ee
\begin{align}
 \cL_{n}(q_\tn{gy}, \dot q_\tn{gy}) &:= - \frac{1}{2} \sum_{l=1}^{n} 
\dot \varrho_{n+1-l} \cdot (\varrho_l \times B_0) - \frac{1}{2} \sum_{l=1}^{n-1} 
\dot \varrho_{n-l} \cdot (\varrho_l \times B_1) + \frac{1}{2} \sum_{l=1}^{n-1} \varrho_{n-l} 
\cdot \nabla E \cdot \varrho_{l} \,\dot t \nonumber
 \\
 &\:\, - \sum_{m=1}^{n} \Big[ \dot G^\para_{n-m} 
(\varrho_m \cdot b_0) + \dot G^\perp_{n-m} (\varrho_m \cdot c_0) - G^\perp_{n-m}\,\dot\alpha\, 
(\varrho_m \cdot 
a_0)  \Big]    \label{def:cLgy}
 \\
 &\:\, +  \sum_{l=1}^{n-1} \Big[\sum_{m=1}^{n-l} G^\para_{n-l-m}\, \dot \varrho_m  \Big] \cdot  
\Big[ 
\varrho_l \cdot 
\nabla b_0 + \cR_{l-1}(b_0) \Big]  \nonumber
 \\
 &\:\, + \sum_{l=1}^{n-1} \Big[ \sum_{m=1}^{n-l} G^\perp_{n-l-m}\, 
 \dot \varrho_m  \Big] \cdot  \Big[ \varrho_l \cdot \nabla c_0 - G^\theta_l\,a_0 + 
\cR_{l-1}^\alpha(c_0) 
\Big]  \nonumber
 \\
 &\:\, + \cR_{n}^\tn{I}(A_0) \cdot \dot r + \sum_{l=2}^{n} \dot \varrho_{n+1-l} \cdot 
\cR_{l-1}(A_0) + \cR_{n-1}^\tn{I}(A_1 \cdot \dot r - \phi\,\dot t) + \sum_{l=2}^{n-1} \dot 
\varrho_{n-l} 
\cdot \cR_{l-1}(A_1) \,. \nonumber
\end{align}
Here, the expressions for $\cR$, $\cR^\alpha$ and $\cR^\tn{I}$ are given in \eqref{def:Rj-1}, 
\eqref{def:Rtheta} and \eqref{def:RjI}, respectively.

\subsection{Proof of Theorem \ref{thm0}}  \label{proof:thm0}

The Lagrangian \eqref{thm0:result} can be written as
\be \label{target}
 L_\tn{gy}^{(N)} = \frac{1}{\eps} L_{-1}^* + L_{0}^* + \sum_{n=1}^N \eps^n L_n^*\,,
\ee
with
$$
 L_{-1}^* = A_0 \cdot \dot r\,,\qquad\quad
 L_{0}^* = (q_\para b_0 + A_1 ) \cdot \dot r - \Big( \frac{q_\para^2}{2} + 
\frac{q_\perp^2}{2} + \phi \Big)\,\dot t \,,\qquad\quad
 L_n^* = \Gamma_n\, \dot \alpha \,,
$$
and $\Gamma_1 = \mu = q_\perp^2/(2|B_0|)$. On the other hand, the series expansion of the 
Lagrangian $L^\eps$ in 
Proposition \ref{prop:1} is composed of the terms
$$
 L_{-1} = A_0 \cdot \dot r\,,\qquad\quad
 L_0 = ( q_\para b_0 + q_\perp c_0 - \varrho_1 \times B_0 + A_1  ) 
\cdot \dot r - \Big( \frac{q_\para^2}{2} + \frac{q_\perp^2}{2} + \phi \Big)\,\dot t \,,
$$
and
\be \label{letsgo}
\begin{aligned}
L_{1\leq n\leq N} &= \Big[ G^\para_{n} b_0 + G^\perp_{n}\, c_0 - \varrho_{n+1} \times 
B_0 - \varrho_n \times B_1 + \cQ_n \Big] \cdot \dot r
 \\[2mm]
 &\quad - ( q_\para G^\para_{n} + q_\perp 
G^\perp_{n} - \varrho_n \cdot E )\,\dot t + \cL_n \,,
\end{aligned}
\ee
where $\cQ_n$ and $\cL_n$ are given in \eqref{def:cQgy} and \eqref{def:cLgy}, respectively.
We shall show that generators can be chosen such that $L_n \sim L_n^*$ for $-1 \leq n\leq N$.
At lowest order one has $L_{-1} = L_{-1}^*$ and nothing needs to 
be done. At zeroth order we choose
 \be \label{u1}
 \varrho_1 = \frac{q_\perp}{|B_0|} b_0 \times c_0 = \frac{q_\perp}{|B_0|} a_0\,,
\ee
which yields $L_0 = L_{0}^*$. For the higher orders we proof the following:

\begin{lemma} \label{lem:9}
 For $n \geq 1$ one can choose generators $\bfG_n,\varrho_{n+1,\perp}$ in the Lagrangian 
\eqref{letsgo}, where $\varrho_{n+1} = (b_0 \cdot \varrho_{n+1})\,b_0 + 
\varrho_{n+1,\perp}$, such that $L_n \sim L_n^* = \Gamma_n\,\dot \alpha$ for arbitrary functions 
$b_0 \cdot \varrho_{n+1}$ and $G_n^\theta$. Moreover, ${\Gamma_1 = \mu = 
q_\perp^2/(2|B_0|)}$ and $\widetilde \Gamma_n = 0$ for all $n$.
\begin{proof}
 We proceed by induction. For $n=1$ we have
 \begin{align}
 \cQ_1 &= \frac{1}{2} 
(\varrho_1 \cdot \nabla B_0) \times \varrho_1 - q_\para\, \varrho_1 
\times (\nabla \times b_0) - q_\perp\, \varrho_1 \times (\nabla \times c_0) - 
 q_\perp\,G^\theta_1\,a_0 \,,  \label{Q1}
 \\[1mm]
 \cL_1 &= - \frac{1}{2}\dot \varrho_1 \cdot (\varrho_1 \times B_0) - 
\dot q_\para\, 
(\varrho_1 \cdot b_0) - \dot q_\perp\, (\varrho_1 \cdot c_0) + q_\perp\,\dot\alpha\, 
(\varrho_1 \cdot a_0)\,.  \label{L1}
\end{align}
Now, from the result \eqref{u1} for $\varrho_1$ we compute 
\be \label{dotu1}
 \dot \varrho_1 = \frac{\dot q_\perp}{|B_0|} a_0 - \frac{q_\perp}{|B_0|^2} (\dot r \cdot \nabla 
|B_0|)\,a_0 + \frac{q_\perp}{|B_0|} \dot r \cdot \nabla a_0 + \frac{q_\perp}{|B_0|}\,\dot 
\alpha\,c_0\,.
\ee
Moreover, from $a_0 \times b_0 = c_0$,
\be \label{id:dot}
 \dot \varrho_1 \cdot (\varrho_1 \times B_0 ) = q_\perp\,\dot \varrho_1 \cdot c_0 =  
\frac{q_\perp^2}{|B_0|} \dot r \cdot \nabla a_0 \cdot c_0 + \frac{q_\perp^2}{|B_0|}\,\dot 
\alpha\,,
\ee
where we recognize the gyro-gauge $\tn R= \nabla a_0 \cdot c_0 = \nabla e_2 \cdot e_1$ .
Inserting this into $\cL_1$  gives
\begin{align}
 L_{1} &= \Big( G^\para_1 b_0 + G^\perp_1\, c_0 - \varrho_2 \times 
B_0 - \frac{q_\perp}{|B_0|} a_0 \times B_1 + \cQ_1 - \frac{q_\perp^2}{2|B_0|}\, \tn R \Big) \cdot 
\dot r    \nonumber
 \\[0mm]
 & \quad - \Big( q_\para G^\para_1 + 
q_\perp G^\perp_1 - \frac{q_\perp}{|B_0|} a_0 \cdot E \Big)\,\dot t  + \frac{q_\perp^2}{2|B_0|} \dot 
\alpha \,.   \label{L1simB}
\end{align}
We can eliminate the Hamiltonian multiplying $\dot t$ by setting
\be \label{h1}
 G^\perp_1 = - \frac{q_\para}{q_\perp} G^\para_1 + \frac{1}{|B_0|}a_0 \cdot E  \,,
\ee
and write the remainder of $L_1$ as
\be \label{L1:remain}
 L_1 = (G^\para_1 b_0 - \varrho_2 \times B_0 + \gamma_1) \cdot \dot r + \frac{q_\perp^2}{2|B_0|} 
\dot \alpha \,.
\ee
Any vector $v\in\RR^3$ can be written as $v = v_\para\,b_0 + v_\perp$, where $v_\para = v \cdot 
b_0$ and $v_\perp = b_0 \times v \times b_0$; hence by setting
\be \label{set:1:zero}
 \varrho_{2,\perp} = \frac{b_0 \times \gamma_1}{|B_0|}\,,\qquad\quad G_1^\para = - 
\gamma_{1,\para}\,,
\ee
we obtain
\be \label{L1:schluss}
 L_1 = \frac{q_\perp^2}{2|B_0|} \dot \alpha = \Gamma_1 \,\dot \alpha  = L_1^* \,.
\ee
Moreover, $b_0 \cdot \varrho_2$ and $G_1^\theta$ are still arbitrary; thus we proved that the 
statement of the lemma holds for $n=1$.

Suppose now that the statement holds for some $n\geq 1$. From Proposition \ref{prop:1} we write the 
Lagrangian at order $n+1$ as
\be \label{L:n+1}
\begin{aligned}
L_{n+1} &= \Big[ G^\para_{n+1} b_0 + G^\perp_{n+1}\, c_0 - \varrho_{n+2} \times 
B_0 - \varrho_{n+1} \times B_1 + \cQ_{n+1} \Big] \cdot \dot r
 \\[2mm]
 &\quad - ( q_\para G^\para_{n+1} + q_\perp 
G^\perp_{n+1} - \varrho_{n+1} \cdot E )\,\dot t + \cL_{n+1} \,,
\end{aligned}
\ee
Let us examine the term $\cL_{n+1}$ from \eqref{def:cLgy} a bit more careful; in particular, let 
us single out two terms:
\begin{itemize}
 \item the term $- \dot q_\para\,(\varrho_{n+1} \cdot b_0)$ from the second line ($m= n+1$),
 \item the term $- q_\perp\, G^\theta_n\,(\dot \varrho_1 \cdot a_0)$ from the fourth line ($l = n$).
\end{itemize}
From \eqref{dotu1} we obtain 
$$
\dot \varrho_1 \cdot a_0 = \frac{\dot q_\perp}{|B_0|} - \frac{q_\perp}{|B_0|^2} (\dot r \cdot 
\nabla |B_0|) \,.
$$
Therefore,
$$
 \cL_{n+1} = - \dot q_\para\,(\varrho_{n+1} \cdot b_0) - \frac{q_\perp}{|B_0|}\, 
G^\theta_n\,\dot q_\perp + \tn{terms}\,.
$$
Moreover, let us add to the Lagrangian $L_{n+1}$ the ``total time derivative'' of some arbitrary 
function $S_{n+1}:\Omega_\tn{gy} \to \RR$, and let us write it in compact notation similarly to 
\eqref{L1:remain}, 
\be \label{L:n+1:B}
\begin{aligned}
L_{n+1} &\sim \Big[ G^\para_{n+1} b_0 - \varrho_{n+2} \times 
B_0 + \gamma_{n+1,r} \Big] \cdot \dot r - \Big( q_\perp 
G^\perp_{n+1} + \gamma_{n+1,t} \Big)\,\dot t
 \\[2mm]
 &\:\: -\Big( \varrho_{n+1} \cdot b_0 + \gamma_{n+1,\para}  \Big)\,\dot q_\para  - 
\Big( \frac{q_\perp}{|B_0|}\,G^\theta_n + \gamma_{n+1,\perp}  \Big)\,\dot q_\perp + 
 \Big( \parfra{S_{n+1}}{\alpha} + \gamma_{n+1,\alpha}\Big) \,\dot \alpha  \,,
\end{aligned}
\ee
where all remaining terms have been gathered in the linear form $\gamma_{n+1}$. Let us treat each 
component of the Poincar\'e-Cartan form on the right-hand-side of \eqref{L:n+1:B} separately:
\begin{itemize}
 \item The component of $\dot r$ is zero for
 $$
 \varrho_{n+2,\perp} = \frac{b_0 \times \gamma_{n+1}}{|B_0|}\,,\qquad\quad G_{n+1}^\para = - 
\gamma_{1,\para}\,.
 $$
 \item The component of $\dot t$ (i.e. the Hamiltonian) is zero for
 $$
  G^\perp_{n+1} = -\frac{\gamma_{n+1,t}}{q_\perp}\,.
 $$
 \item Since $b_0 \cdot \varrho_{n+1} $ is still undetermined by the inductive hypothesis, the 
component of 
$\dot q_\para$ is zero for
 $$
  b_0 \cdot \varrho_{n+1} = -\gamma_{n+1,\para}\,.
 $$
 \item Noting that $G_n^\theta$ is still undetermined by the inductive hypothesis, the component of 
$\dot q_\perp$ is zero for
 $$
  G_n^\theta = -\frac{|B_0|}{q_\perp}\,\gamma_{n+1,\perp}\,.
 $$
 \item The term with $\dot \alpha$ is rewritten as
 $$
  \Big( \parfra{S_{n+1}}{\alpha} + \gamma_{n+1,\alpha}\Big) \,\dot \alpha = \Big( 
\parfra{S_{n+1}}{\alpha} + \gavg{\gamma_{n+1,\alpha}} + \widetilde{\gamma_{n+1,\alpha}}\Big) \,\dot 
\alpha \,,
 $$
 where $\gamma_{n+1,\alpha}$ has been decomposed into gyro-average and fluctuations. The equation
 $$
 \parfra{S_{n+1}}{\alpha} + \widetilde{\gamma_{n+1,\alpha}} = 0
 $$
 has $2\pi$-periodic solutions $S_{n+1}$. We pick one of those solutions to obtain
 $$
 \Big( \parfra{S_{n+1}}{\alpha} + \gamma_{n+1,\alpha}\Big) \,\dot \alpha = 
\gavg{\gamma_{n+1,\alpha}} \,\dot \alpha\,.
 $$
 \end{itemize}
 Hence, with the above choices for the generators, all that remains from \eqref{L:n+1:B} is
 $$
 L_{n+1} \sim \gavg{\gamma_{n+1,\alpha}} \,\dot \alpha =: \Gamma_{n+1}\,\dot \alpha = L_{n+1}^*\,.
 $$
Noting that $b_0 \cdot \varrho_{n+2}$ and $G_{n+1}^\theta$ are still arbitrary and that 
$\widetilde \Gamma_{n+1} = 0$, we showed that the statement of the lemma holds for $n+1$ and 
thus completed the proof by induction.
\end{proof}
\end{lemma}

Considering the regularity of the generators $\bfG_n$ the following is true:

\begin{lemma}  \label{lem:10}
 For $1 \leq n\leq N$ we have $\bfG_n \in C^{N+2-n}(\Omega_\tn{gy})$ and $\varrho_{n+1,\perp} \in 
C^{N+2-n}(\Omega_\tn{gy})$.
 \begin{proof}
  The proof is again achieved by induction. For $n=1$ the generators 
$\varrho_1,G^\para_1,G^\perp_1$ and $\varrho_{2,\perp}$ are given in Lemma \ref{lemma:2} and 
$G^\theta_1$ is given in Lemma \ref{lemma:2b}, respectively. From assumption \ref{assumps} we 
deduce $\bfG_1,\varrho_{2,\perp} \in C^{N+1}(\Omega_\tn{gy})$. Assuming the statement holds 
for some $n\leq N-1$, it follows from the proof of Lemma \ref{lem:9} that the generators 
$\varrho_{n+2,\perp},G^\para_{n+1},G^\perp_{n+1}$ and $b_0 \cdot \varrho_{n+1}$ have the same 
regularity as the Lagrangian $L_{n+1}$ written in \eqref{L:n+1}. The fact that $L_{n+1} \in 
C^{N+2-(n+1)}$ follows from $\dot \varrho_{n+1,\perp} \in C^{N+2-n-1}$ due to the inductive 
hypothesis (needed in the first term of $\cL_{n+1}$, equation \eqref{def:cLgy}) as well as from 
$\cR_n(b_0),\cR^\alpha_n(c_0),\cR^\tn{I}_{n+1}(A_0)$ all being in $C^{N+2-(n+1)}$.

It remains to determine the regularity of the gererator $G_{n+1}^\theta$, which is the same as the 
one of the terms multplying $\dot q_\perp$ in $\cL_{n+2}$, according to the proof of Lemma 
\ref{lem:9}. A close inspection of \eqref{def:cLgy} reveals that such terms can only stem from  
$\dot \varrho_{n+2,\perp}$, $\dot \varrho_{n+1}$, $\dot G^\para_{n+1}$ and $\dot G^\perp_{n+1}$. But 
derivation with respect to $q_\perp$ does not change the regularity  since everything is $C^\infty$ 
in the velocities; therefore, $G_{n+1}^\theta \in C^{N+2-(n+1)}$ and the proof is complete.
 \end{proof}
\end{lemma}

Taking the statement from Lemma \ref{lem:10} for $n=N$ we have $\bfG_N \in 
C^{2}(\Omega_\tn{gy})$ and on the next level $\varrho_{N+1,\perp} \in C^{2}(\Omega_\tn{gy})$. 
According to Lemma \ref{lem:9} all other generators at the level $N+1$ can be set to zero and thus
Theorem \ref{thm0} is proved.
\qed

\subsection{Proof of Lemma \ref{lem:2}}  \label{proof:lem:2}

\begin{lemma} \label{Gronwall}
 ({\bf Gronwall} \cite{Sanders2007}) Suppose that for $t\in[t_0,t_0+T]$
 \bes
 \vphi(t) \leq b\,(t-t_0) + a \int_{t_0}^t \vphi(s) ds + c\,,
 \ees
 with $\vphi(t)$ continuous, $\vphi(t)\geq 0$ for $t\in[t_0,t_0+T]$ and constants $a>0$, $b,c \geq 
0$, then
 \bes
 \vphi(t) \leq \Big( \frac{b}{a} + c \Big) e^{a(t-t_0)} - \frac{b}{a}
 \ees
 for $t\in[t_0,t_0+T]$.
\end{lemma}

In order to set the framework necessary to apply Gronwall's lemma, let us write \eqref{Pprobs:sub} 
as integral equations,
  \begin{align*}
   \bfz(t) &= \bfz_0 + \int_{t_0}^t [ \Lambda(\bfz,s,\eps) + \eps^{N} 
S(\bfz,s,\eps)] ds\,,
   \\[0mm]
   \overline \bfz(t) &= \overline \bfz_0 + \int_{t_0}^t \Lambda(\overline 
\bfz,s,\eps)  ds\,.
  \end{align*}
  Subtracting the equations and taking the norm yields
  \begin{align*}
  &|| \overline \bfz(t) -  \bfz(t) || = || \overline \bfz_0 -  
\bfz_0 + \int_{t_0}^t [ \Lambda(\overline\bfz,s,\eps) - \Lambda( 
\bfz,s,\eps) - \eps^{N} 
S(\bfz,\alpha,s,\eps)] ds \,||
 \\[0mm]
 &\qquad\leq || \overline \bfz_0 - \bfz_0|| +  \int_{t_0}^t || 
\Lambda( \overline \bfz,s,\eps) - \Lambda( 
\bfz,s,\eps) ||ds + \eps^N 
\int_{t_0}^t 
||S(\bfz,\alpha,s,\eps)|| ds \,.
  \end{align*}
  The residual $S$ is continuous; a solution $\bfz(s)$ of $(P^\eps)$ is too (remark 
\ref{lem:3}) and lives in a bounded domain with size $O(1)$ as $\eps\to 0$. Therefore, 
$$
||S(\bfz(s),\alpha,s,\eps)|| \leq ||S||_{\infty,\eps} := \max_{\wh \Omega_\tn{gy} \times 
(0,\eps_\tn{max})} 
||S(\bfz,\alpha,s,\eps)|| = O(1)\,.
$$
Moreover, $\Lambda$ is Lipschitz with constant $\ell_\Lambda$ and we 
may estimate
 \bes
 || \overline \bfz(t) - \bfz(t) || \leq || \overline \bfz_0 -  
\bfz_0|| +  \ell_\Lambda \int_{t_0}^t || \overline  \bfz(s) - \bfz(s) ||ds + 
\eps^N 
(t-t_0) ||S||_\infty \,.
 \ees
 We now apply Gronwall's lemma with $\vphi(t) = ||\overline \bfz(t) -  
\bfz(t) ||$, $a =  
\ell_\Lambda$, $b = \eps^N ||S||_{\infty,\eps}$ and $c = ||  \overline \bfz_0 - \bfz_0||$ to 
obtain
$$
  || \overline \bfz(t) -  \bfz(t) ||  \leq \Big( \eps^N \frac{||S||_{\infty,\eps}}{\ell_\Lambda} 
+ || \overline \bfz_0 -  \bfz_0|| \Big) 
e^{ 
\ell_\Lambda(t-t_0)} - \eps^N \frac{||S||_{\infty,\eps}}{\ell_\Lambda}\,.
$$
\qed

\section{Conclusion} \label{sec:concl}

Gyrokinetics is a prevalent theory in plasma physics; it enables the numerical simulation of 
sophisticated multiscale physics on long timescales. The contribution of this work is to build a 
mathematically sound foundation for gyrokinetics by means of averaging systems of 
differential equations on the level of the Lagrangian function, hence the name variational 
averaging (VA). The formal theory of VA is well-known for almost three decades; the most important 
results with emphasis on applications in plasma physics have been gathered in recent 
reviews \cite{Brizard_rev2007,Krommes2012gyrokinetic}. The theory has here been made rigorous in 
the following sense:
\begin{itemize}
 \item The theory starts from the normalized set of equations \eqref{ivp} and stays 
consistent with this scaling thorough all orders of the perturbation expansion.
 \item The gyro-transformations \eqref{transf2} employed in this work exist, c.f. Theorem 
\ref{thm0}. This is in contrast to the formal theories, where transformations are infinite series 
of which the convergence cannot be established.
 \item We state the unambiguous definition of a gyrokinetic equation in \eqref{def:GK} by means of 
the ``decoupled dynamics'' \eqref{dt:avg}, which stem from a truncated Lagrangian function. 
 \item For the first time we give an error estimate for gyrokinetics in Theorem \ref{thm}.
\end{itemize}
The method of VA is thus well-established for the charged particle motion. It seems plausible that 
this technique could be applied also to other problems of averaging, where the Hamiltonian 
structure of the equations is important and which are non-canonical symplectic, i.e.
with a Lagrangian of the form \eqref{intro:L}. For instance, an application of VA to the 
semi-classical limit of the Schr\"odinger equation could be envisioned. But also other fields like 
liquid crystal dynamics might be an interesting possibility for application of the VA-method. 
Moreover, the relation between VA and other averaging methods, in particular normal forms, should 
be clarified. 

Regarding the charged particle, let us comment on some of the practical implications of the here 
derived results. In view of the GY-Lagrangian from Theorem \ref{thm0}, repeated in equation 
\eqref{L:trunc} with the generalized magnetic moment $\wh\mu$ as one of the coordinates, we remark 
that only the Hamiltonian $H_\tn{gy}$ depends on the order $N$ of the perturbation expansion, 
whereas the symplectic form remains unchanged through all orders. This is remarkable because we did 
not make any particular effort to achieve this; in conventional GY-theories this is usually enforced 
by an ansatz for the GY-transformation in the form of a Lie-series. Here, the formalism is 
considerably simpler.

Expressions for the Hamiltonian $H_\tn{gy}$ have been computed for $N=2$ in the Lemmas 
\ref{lemma:4b} and \ref{lemma:2b}. They differ from the conventional GY-Hamiltonians as was pointed 
out in Remark~\ref{rem:5}. This is not a surprise considering the amount of freedom 
within the VA methodology: at each order $n$, there is a choice to be made which terms of the 
Lagrangian $L_n$ in the series \eqref{prop:1:result} should be attached to the generators, and thus 
appear in the transformation, and which should be kept in the Lagrangian, and thus appear in the 
dynamics. In conventional GY-theories the gyro-average of the Lagrangian constitutes the dynamics, 
while the fluctuating part disappears into the generators. However, this is not mandatory. Our 
approach was to attach as many terms as possible to the generators, even gyro-averaged terms, 
thereby keeping the dynamics simpler. This could be beneficial for a certain class of numerical 
codes, in particular particle-based codes, in which an efficient particle pusher is important. We 
plan the implementation of such a scheme in a forthcoming work. 

Finally, the error estimate in Theorem \ref{thm} relies on the the assumption  
that the gyrokinetic initial condition has gyro-fluctuations of the order $O(\eps^N)$; this is 
called a ``well-prepared'' initial condition. In the estimate we compare the solution of the 
averaged part \eqref{GK:avg} of the gyrokinetic equation to the solution of the Vlasov equation, 
transformed to the new coordinates, which depends on the gyro-angle $\alpha$. It is thus clear 
that the error is small only when the $\alpha$-dependence of the Vlasov solution $f$ is. In 
practice one is often faced with the computation of velocity moments 
of $f$, which is why we chose to focus on the estimate from Theorem \ref{thm}.

\section{Acknowledgements} \label{sec:acknowl}

I am thankful to Eric Sonnendr\"ucker for encouraging this work and for his valuable scientific 
input. I also thank Francis Filbet, Michael Kraus and Edoardo Zoni for the inspiring discussion I 
had with them as well as Roman Hatzky for reading the manuscript with care.


\begin{thebibliography}{10}

\bibitem{Marsden}
R.~Abraham and J.E. Marsden.
\newblock {\em {Foundations of Mechanics}}.
\newblock Addison-Wesley, 1978.

\bibitem{Arnold}
V.I. Arnold.
\newblock {\em {Mathematical Methods of Classical Mechanics}}.
\newblock Number~60 in {Graduate Texts in Mathematics}. Springer, 2nd edition,
  1989.

\bibitem{Bostan2007}
M. Bostan.
\newblock {The Vlasov--Maxwell system with strong initial magnetic field:
  guiding-center approximation}.
\newblock {\em Multiscale Modeling \& Simulation}, 6(3):1026--1058, 2007.

\bibitem{Bostan2010}
M. Bostan.
\newblock {Gyrokinetic Vlasov equation in three dimensional setting. Second
  order approximation}.
\newblock {\em Multiscale Modeling \& Simulation}, 8(5):1923--1957, 2010.

\bibitem{Bostan2010transport}
M. Bostan.
\newblock {Transport equations with disparate advection fields. Application to
  the gyrokinetic models in plasma physics}.
\newblock {\em Journal of Differential Equations}, 249(7):1620--1663, 2010.

\bibitem{Bostan2016}
M. Bostan.
\newblock {MultiScale Analysis for Linear First Order PDEs. The Finite Larmor
  Radius Regime }.
\newblock {\em SIAM Journal on Mathematical Analysis}, 48(3):2133--2188, 2016.

\bibitem{ORB5}
A. Bottino, B. Scott, S. Brunner, B.~F. McMillan, T.~M. Tran,
  T. Vernay, L. Villard, S. Jolliet, R. Hatzky, and
  A.~G. Peeters.
\newblock {Global nonlinear electromagnetic simulations of tokamak turbulence}.
\newblock {\em IEEE Transactions on Plasma Science}, 38(9):2129--2135, 2010.

\bibitem{Brizard1989}
A.~Brizard.
\newblock {Nonlinear gyrokinetic Maxwell-Vlasov equations using magnetic
  co-ordinates}.
\newblock {\em Journal of plasma physics}, 41(3):541--559, 1989.

\bibitem{Brizard_rev2007}
A.J. Brizard and T.S. Hahm.
\newblock {Foundations of nonlinear gyrokinetic theory}.
\newblock {\em Rev. Mod. Phys.}, 79:421, 2007.

\bibitem{Burby}
J.W. Burby, J.~Squire, and H.~Qin.
\newblock {Automation of the guiding center expansion}.
\newblock {\em Physics of Plasmas}, 20:072105, 2013.

\bibitem{GYRO}
J.~Candy and R.E.~Waltz.
\newblock {Anomalous transport scaling in the DIII-D tokamak matched by
  supercomputer simulation}.
\newblock {\em Physical review letters}, 91(4):045001, 2003.

\bibitem{BrizardCary2009}
J.~R. Cary and A.~J. Brizard.
\newblock {Hamiltonian theory of guiding-center motion}.
\newblock {\em Reviews of modern physics}, 81(2):693, 2009.

\bibitem{Chartier2016}
P. Chartier, N. Crouseilles, and M. Lemou.
\newblock {An averaging technique for transport equations}.
\newblock {\em arXiv preprint arXiv:1609.09819}, 2016.

\bibitem{Chartier2012}
P. Chartier, A. Murua, and J.~M. Sanz-Serna.
\newblock {A formal series approach to averaging: exponentially small error
  estimates}.
\newblock {\em Discrete and Continuous Dynamical Systems-Series A}, 32(9),
  2012.

\bibitem{Filbet2016asymp}
P. Degond and F. Filbet.
\newblock {On the Asymptotic Limit of the Three Dimensional Vlasov--Poisson
  System for Large Magnetic Field: Formal Derivation}.
\newblock {\em Journal of Statistical Physics}, 165(4):765--784, 2016.

\bibitem{Frankel}
T. Frankel.
\newblock {\em {The geometry of physics: an introduction}}.
\newblock Cambridge University Press, 2011.

\bibitem{FrenodSonnen1998}
E. Fr{\'e}nod and E. Sonnendr{\"u}cker.
\newblock {Homogenization of the Vlasov equation and of the Vlasov--Poisson
  system with a strong external magnetic field}.
\newblock {\em Asymptotic Analysis}, 18(3-4):193--213, 1998.

\bibitem{FrenodSonnen2001}
E. Fr{\'e}nod and E. Sonnendr{\"u}cker.
\newblock {The finite Larmor radius approximation}.
\newblock {\em SIAM Journal on Mathematical Analysis}, 32(6):1227--1247, 2001.

\bibitem{Golse1999}
F. Golse and L. Saint-Raymond.
\newblock {The Vlasov--Poisson system with strong magnetic field}.
\newblock {\em Journal de math{\'e}matiques pures et appliqu{\'e}es},
  78(8):791--817, 1999.

\bibitem{GENE}
T.~G{\"o}rler, X.~Lapillonne, S. Brunner, T. Dannert, F. Jenko,
  F. Merz, and D.~Told.
\newblock {The global version of the gyrokinetic turbulence code GENE}.
\newblock {\em Journal of Computational Physics}, 230(18):7053--7071, 2011.

\bibitem{GYSELA}
V.~Grandgirard, Y.~Sarazin, X.~Garbet, G.~Dif-Pradalier, Ph. Ghendrih,
  N.~Crouseilles, G.~Latu, E.~Sonnendr{\"u}cker, N.~Besse, and P.~Bertrand.
\newblock {GYSELA, a full-f global gyrokinetic Semi-Lagrangian code for ITG
  turbulence simulations}.
\newblock {\em AIP Conference Proceedings}, 871(1):100--111, 2006.

\bibitem{Hahm1988}
T.S.~Hahm.
\newblock {Nonlinear gyrokinetic equations for tokamak microturbulence}.
\newblock {\em The Physics of fluids}, 31(9):2670--2673, 1988.

\bibitem{Hahm1996}
T.S.~Hahm.
\newblock {Nonlinear gyrokinetic equations for turbulence in core transport
  barriers}.
\newblock {\em Physics of Plasmas}, 3(12):4658--4664, 1996.

\bibitem{Hairer_geom}
E. Hairer, C. Lubich, and G. Wanner.
\newblock {\em {Geometric Numerical Integration}}, volume~31.
\newblock Springer Series in Computational Mathematics, 2006.

\bibitem{Han2010}
D. Han-Kwan.
\newblock {The three-dimensional finite Larmor radius approximation}.
\newblock {\em Asymptotic Analysis}, 66(1):9--33, 2010.

\bibitem{Hazeltine}
R.~D. Hazeltine and J.~D. Meiss.
\newblock {\em {Plasma Confinement}}.
\newblock Dover Books, 2003.

\bibitem{ELMFIRE}
J.A.~Heikkinen, S.~Henriksson, S.~Janhunen, T.P.~Kiviniemi, and F.~Ogando.
\newblock {Gyrokinetic simulation of particle and heat transport in the
  presence of wide orbits and strong profile variations in the edge plasma}.
\newblock {\em Contributions to Plasma Physics}, 46(7-9):490--495, 2006.

\bibitem{Jackson}
J.~D. Jackson.
\newblock {\em {Classical electrodynamics}}.
\newblock John Wiley \& Sons, 2007.

\bibitem{EUTERPE}
G.~Jost, T.M.~Tran, W.A.~Cooper, L.~Villard, and K.~Appert.
\newblock {Global linear gyrokinetic simulations in quasi-symmetric
  configurations}.
\newblock {\em Physics of Plasmas}, 8(7):3321--3333, 2001.

\bibitem{Krommes2012gyrokinetic}
J.~A. Krommes.
\newblock {The gyrokinetic description of microturbulence in magnetized
  plasmas}.
\newblock {\em Annual Review of Fluid Mechanics}, 44:175--201, 2012.

\bibitem{Kruskal}
M.~Kruskal.
\newblock {Asymptotic Theory of Hamiltonian and other Systems with all
  Solutions Nearly Periodic}.
\newblock {\em J. Math. Phys.}, 3:806, 1962.

\bibitem{GTC}
Z. Lin, T.~S. Hahm, W.W.~Lee, W.~M. Tang, and R.~B. White.
\newblock {Turbulent transport reduction by zonal flows: Massively parallel
  simulations}.
\newblock {\em Science}, 281(5384):1835--1837, 1998.

\bibitem{Little1979}
R.G. Littlejohn.
\newblock {A guiding center Hamiltonian: A new approach}.
\newblock {\em J. Math. Phys.}, 20(12), 1979.

\bibitem{Little1983}
R.G. Littlejohn.
\newblock {Variational principles of guiding center motion}.
\newblock {\em J. Plasma Physics}, 29:111--125, 1983.

\bibitem{Little1981}
R.~G. Littlejohn.
\newblock {Hamiltonian formulation of guiding center motion}.
\newblock {\em The Physics of Fluids}, 24(9):1730--1749, 1981.

\bibitem{Little1982}
R.~G. Littlejohn.
\newblock {Hamiltonian perturbation theory in noncanonical coordinates}.
\newblock {\em Journal of Mathematical Physics}, 23(5):742--747, 1982.

\bibitem{Northrop1963}
T.~G. Northrop.
\newblock {Adiabatic charged-particle motion}.
\newblock {\em Reviews of Geophysics}, 1(3):283--304, 1963.

\bibitem{GEM}
S.E.~Parker, Y.~Chen, W.~Wan, B.I.~Cohen, and W.M.~Nevins.
\newblock {Electromagnetic gyrokinetic simulations}.
\newblock {\em Physics of Plasmas}, 11(5):2594--2599, 2004.

\bibitem{Parra2011}
F.~I. Parra and I. Calvo.
\newblock {Phase-space Lagrangian derivation of electrostatic gyrokinetics in
  general geometry}.
\newblock {\em Plasma Physics and Controlled Fusion}, 53(4):045001, 2011.

\bibitem{Sanders2007}
J.~A. Sanders, F. Verhulst, and J.~A. Murdock.
\newblock {\em {Averaging methods in nonlinear dynamical systems}}, volume~59.
\newblock Springer, 2007.

\bibitem{Scott2017}
B.~D. Scott.
\newblock {Gyrokinetic Field Theory as a Gauge Transform or: gyrokinetic theory
  without Lie transforms}.
\newblock {\em arXiv preprint arXiv:1708.06265}, 2017.

\bibitem{Tronci2016}
C. Tronci.
\newblock {From liquid crystal models to the guiding-center theory of
  magnetized plasmas}.
\newblock {\em Annals of Physics}, 371:323--337, 2016.

\bibitem{Brizard_Natalia2015}
N.~Tronko and A.J. Brizard.
\newblock {Lagrangian and Hamiltonian constraints for guiding-center
  Hamiltonian theories}.
\newblock {\em Physics of Plasmas}, 22(11):112507, 2015.

\bibitem{Tronko2017}
N. Tronko, A. Bottino, C. Chandre, and E. Sonnendruecker.
\newblock {Hierarchy of second order gyrokinetic Hamiltonian models for
  particle-in-cell codes}.
\newblock {\em Plasma Physics and Controlled Fusion}, 59(6):064008, 2017.

\end{thebibliography}
\bibliographystyle{plain}

\end{document}